\documentclass[aps,pre]{revtex4-1}
\usepackage{amsmath,amssymb,latexsym,epsfig,graphics,epsf}
\usepackage{graphics,xcolor}
\usepackage{float}
\usepackage{hyperref}
\newcommand{\be}{\begin{equation}}
\newcommand{\ee}{\end{equation}}
\newcommand{\fig}[1]{Fig.~\ref{#1}}

\newcommand{\eq}[1]{Eq.~(\ref{#1})}

\begin{document}
	\title{Single-parameter aging in a binary Lennard-Jones system}
	\date{\today}
	\author{Saeed Mehri}\email{mehri@ruc.dk}
	\author{Trond S. Ingebrigtsen}
	\author{Jeppe C. Dyre}\email{dyre@ruc.dk}
	\affiliation{Glass and Time, IMFUFA, Department of Science and Environment, Roskilde University, P.O. Box 260, DK-4000 Roskilde, Denmark}
	
\begin{abstract}
This paper studies physical aging by computer simulations of a 2:1 Kob-Andersen binary Lennard-Jones mixture, a system that is less prone to crystallization than the standard 4:1 composition. Starting from thermal-equilibrium states, the time evolution of the following four quantities is monitored following up and down jumps in temperature: the potential energy, the virial, the average squared force, and the Laplacian of the potential energy. Despite the fact that significantly larger temperature jumps are studied here than in previous experiments, to a good approximation all four quantities conform to the single-parameter-aging scenario derived and validated for small jumps in experiments [Hecksher \textit{et al.}, J. Chem. Phys. \textbf{142}, 241103 (2015)]. As a further confirmation of single-parameter aging with a common material time for the different quantities monitored, their relaxing parts are found to be almost identical for all temperature jumps.
\end{abstract}
\maketitle

\section{Introduction}

It is of great interest to be able to predict how much and how fast material properties change over time \cite{cangialosi2014dynamics}. Such gradual property changes are referred to as aging. Corrosion and weathering in general give rise to aging. The term ``physical aging'' refers to changes of material properties that result exclusively from molecular rearrangements, i.e., involve no chemical changes \cite{Tool1931,Narayana1971,scherer1986relaxation}. A number of theories of physical aging have been developed \cite{Tool1931,Narayana1971,Hodge1945,moynihan1976dependence,scherer1986relaxation,olsen1998structural,Hecksher2015,mauro2009fictive,cugliandolo1994out,kob2000fluctuations,adolf2007potential,castillo2007local,parsaeian2009equilibrium,kolvin2012simple}, and physical aging has been the subject of experimental studies in different contexts dealing, e.g., with oxide glasses \cite{Narayana1971,scherer1986relaxation}, polymers \cite{Struik1978PhysicalAI,Hodge1945,hutchinson1995prog,odegard2011physical,cangialosi2013physical,grassia2012modeling}, metallic glasses \cite{qiao2014dynamic,son20}, colloids \cite{son20}, and spin glasses \cite{lundgren1983dynamics,berthier2002geometrical}. Examples of quantities monitored in order to probe physical aging are: density \cite{cangialosi2013physical,spinner1966further}, enthalpy \cite{Narayana1971,moynihan1976dependence}, Young's modulus \cite{cangialosi2013physical}, and various frequency-dependent responses \cite{schlosser1991dielectric,leheny1998frequency,lunkenheimer2005glassy,richert2015supercooled,olsen1998structural,Hecksher2015,hecksher2010physical,wehn2007broadband,dyre2003minimal}.

Physical aging is generally both non-exponential and non-linear. The latter property is reflected in the fact that the system's response to a small perturbation depends on both the sign and the magnitude of the input. Ideally, an aging experiment consists of an up or a down jump in  temperature starting from a state of thermal equilibrium, eventually ending in equilibrium at the ``target'' (annealing) temperature. The hallmark of aging is that these two responses, even if they go to the same temperature, are \textit{not} mirror symmetric. A down jump is fast at the beginning, but slows down gradually as equilibrium is approached (``self-retarding''). An up jump -- while slower in the beginning -- will after an initial delay show a steeper approach to equilibrium (``self-accelerating'') 
\cite{cangialosi2013physical,mauro2009fictive,scherer1986relaxation}. This is the fictive-temperature effect, also referred to as ``asymmetry of approach'' \cite{kov63,di11,mck17}, an effect that is well understood as a consequence of the fact that the relaxation rate itself ages
\cite{Struik1978PhysicalAI,Hodge1945,moynihan1976dependence,scherer1986relaxation,mazurin1977relaxation,mckenna1994physics,hodge1994enthalpy,avramov1996kinetics}.

In experimental studies of physical aging, the temperature $T$ is externally controlled and identified as the phonon ``bath'' temperature measured on a thermometer. Recently, Hecksher \textit{et al.} \cite{Hecksher2015} and Roed \textit{et al.} \cite{roed2019generalized} studied the physical aging of glass-forming liquids around the glass transition temperature by probing the shear-mechanical resonance frequency ($\sim 360$kHz), the dielectric loss at $1$Hz, the real part of the dielectric constant at $10$kHz, and the loss-peak frequency of the dielectric beta process ($\sim 10$kHz). These authors developed a ``single-parameter aging'' (SPA) formalism as a simple realization of the Narayanaswamy idea that a material time controls aging \cite{Narayana1971}. SPA basically allows one to predict the normalized relaxation functions of an arbitrary temperature jump from the data of a single jump. SPA was first demonstrated for jumps to the same temperature for three different van der Waals liquids \cite{Hecksher2015}, and subsequently generalized to deal with jumps ending at different temperatures in a study of glycerol \cite{roed2019generalized}. 

The motivation of the current study is to illuminate how general SPA is by investigating whether SPA applies also in computer simulations. The advantage of simulations is that one can probe well-defined microscopic quantities and, for instance, easily study the aging of several different quantities under identical circumstances. We report below data for the physical aging of a binary Lennard-Jones mixture upon a temperature jump. The following four quantities were monitored: virial, potential energy, the average squared force, and the Laplacian of the potential energy. All four quantities conform to SPA to a good approximation, even for temperature jumps as large as 10\%.

\section{The Tool-Narayanaswamy material-time concept}\label{sec2}

Above the melting temperature a liquid is rarely particularly viscous. At lower temperatures, the liquid becomes supercooled and, because of the extraordinary large viscosities reached upon further cooling, it gradually behaves more like a ``solid that flows'' than like an ordinary liquid \cite{dyr06}. For both the ordinary liquid phase and the glass phase, under ambient pressure conditions physical properties are found to depend only on the temperature. At temperatures in the vicinity of the glass transition temperature (defined by the applied cooling and heating rates), however, the behavior is different. In this temperature range, the molecular structure changes gradually with temperature, and following an external perturbation a noticeable delay is observed before equilibrium is reached. In this case, the physical properties depend not just on the actual temperature, but on the entire thermal history of the system.

In 1971 Narayanaswamy established what has become the standard formalism for physical aging. It was developed for predicting how the frozen-in stresses in a wind shield depend on the glass' thermal history during production. The theoretical framework, which turned out to be generally applicable for physical aging involving moderate temperature changes \cite{Narayana1971,scherer1986relaxation}, is now referred to as the Tool-Narayanaswamy (TN) formalism. This framework systematically addresses the non-exponential and non-linear nature of aging. The TN formalism reproduces all observed qualitative features of physical aging and it is in most cases in quantitative agreement with experiments \cite{scherer1986relaxation,ritland1956limitations,Narayana1971,Tool1931,dyre2015narayanaswamy}. 

The crucial concept of TN is that of a \textit{material time}, denoted below by $\xi$. The material time may be thought of as the time measured on a clock with a clock rate, $\gamma(t)$, that changes as the material ages. Simply put, the material time is the time that a substance ``experiences'', which in equilibrium is proportional to the actual time. In this physical picture, one expects the existence of a single material time controlling the physical aging of different quantities. 

Since the clock rate by definition measures how fast the material time changes \cite{Narayana1971,dyre2015narayanaswamy}, one has

\be\label{xi}
d\xi = \gamma(t)dt\,.
\ee
Narayanaswamy showed from experimental data that if one switches from time to material time in the description of aging, the aging response becomes linear. In other words, a non-linear aging response is described by a \textit{linear} convolution integral over the material time \cite{Narayana1971,scherer1986relaxation}. This was an important and highly nontrivial finding. For instance, it implies that the ``asymmetry of approach'' becomes a ``symmetry of approach'' when jumps of equal magnitude to the same temperature are considered as functions of the material time.

\section{Simulation details}

The simulations were performed in the $NVT$ ensemble with the Nos\'e-Hoover thermostat using the RUMD (Roskilde University Molecular Dynamics) GPU open-source code [http://rumd.org]. We simulated a system of $10002$ particles consisting of two different Lennard-Jones (LJ) spheres, A and B. Writing the LJ pair potential between particles of type $\alpha$ and $\beta$ as $v_{\alpha \beta}(r)=\varepsilon_{\alpha \beta}((r/\sigma_{\alpha \beta})^{-12}-(r/\sigma_{\alpha \beta})^{-6})$ ($\alpha,\beta =A,B$), the parameters used are $\sigma_{AA}=1.0$, $\sigma_{AB}=\sigma_{BA}=0.8$, $\sigma_{BB}=0.88$, $\varepsilon_{AA}=1.0$, $\varepsilon_{AB}=\varepsilon_{BA}=1.5$, $\varepsilon_{BB}=0.5$. All simulations employed the MD time step $0.0025$ (in the units defined by the A particle parameters) and a shifted-potential cutoff of $v_{\alpha \beta}(r)$ at $r_\textrm{cut}=2.5\sigma_{\alpha \beta}$. The pair-potential parameters are the same as those of the well-known Kob-Andersen (KA) mixture \cite{ka1}, which has previously been used for numerical studies of physical aging and other glass-transition related non-equilibrium phenomena \cite{vol96,par08a,reh10,gna13,pri18}. We use a ratio of A and B particles that is 2:1 instead of the standard 4:1 ratio, however, because the 2:1 mixture is much more resistant toward crystallization than the 4:1 composition \cite{ped18,PhysRevX.9.031016,bel20} (an alternative option for avoiding crystallization of KA mixtures is to keep the 4:1 composition and employ a short-distance shifted-force cutoff for the AA and BB interactions \cite{sch20}). For comparison of results for the two different compositions, we note that the mode-coupling temperature is around $0.55$ for the 2:1 KA system whereas it is around $0.44$ for the standard mixture. 

All results reported below were obtained at the density 1.4 (in A particle units) and represent averaging over $100$ simulations. Before performing a temperature jump, the system was carefully equilibrated. To reach equilibrium at the lowest temperature ($T=0.50$) and ensure that there is no crystallization issue, we first simulated the system by performing $2.4\times 10^{11}$ time steps. After this, each of 100 equilibrium configurations were obtained by dumping a configuration every $1.678\times 10^{8}$ time steps. At higher temperature equilibrium is reached much faster, of course.

\begin{figure}[htbp!]
	\includegraphics[width=7cm]{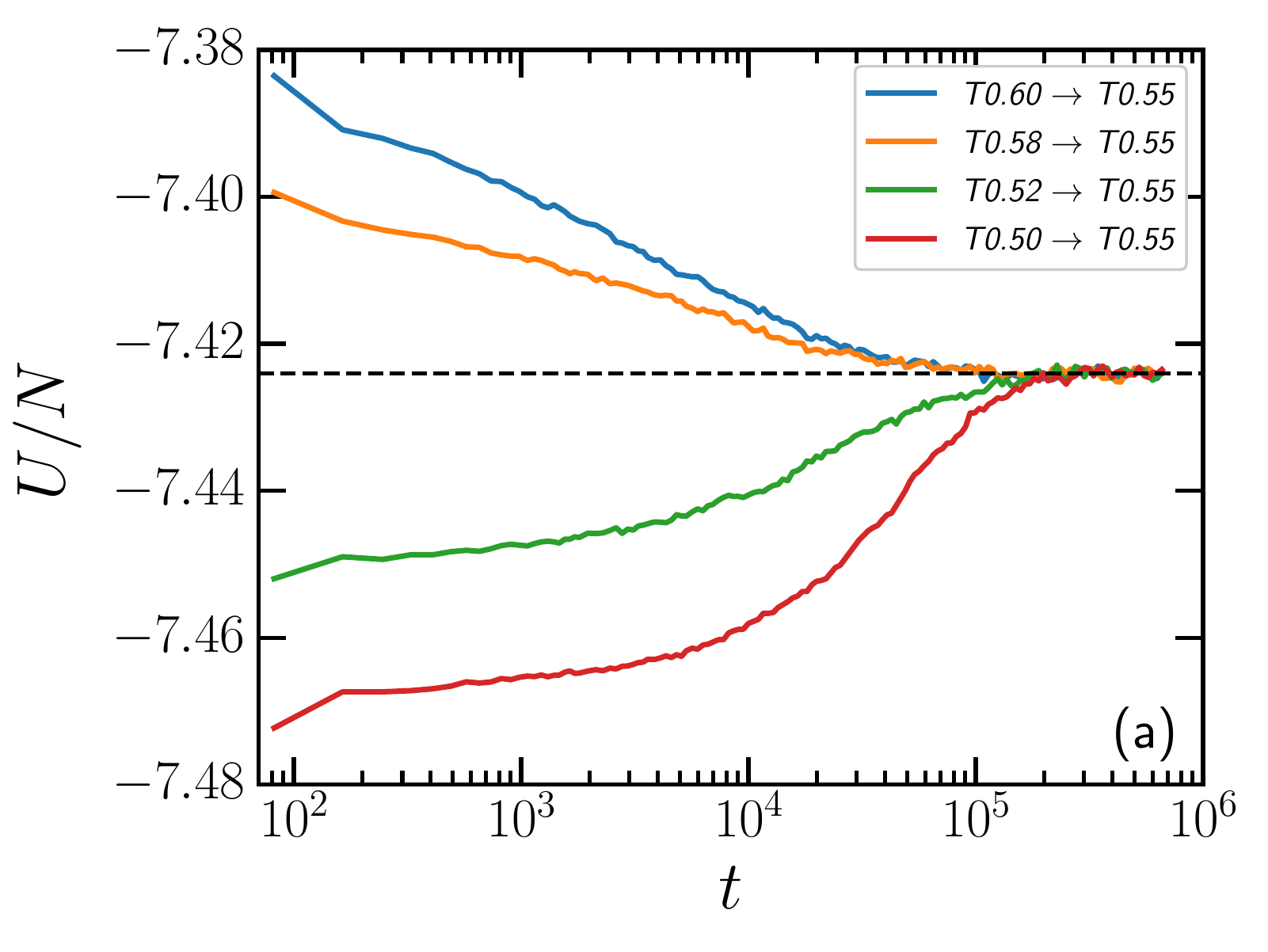}
	\includegraphics[width=7cm]{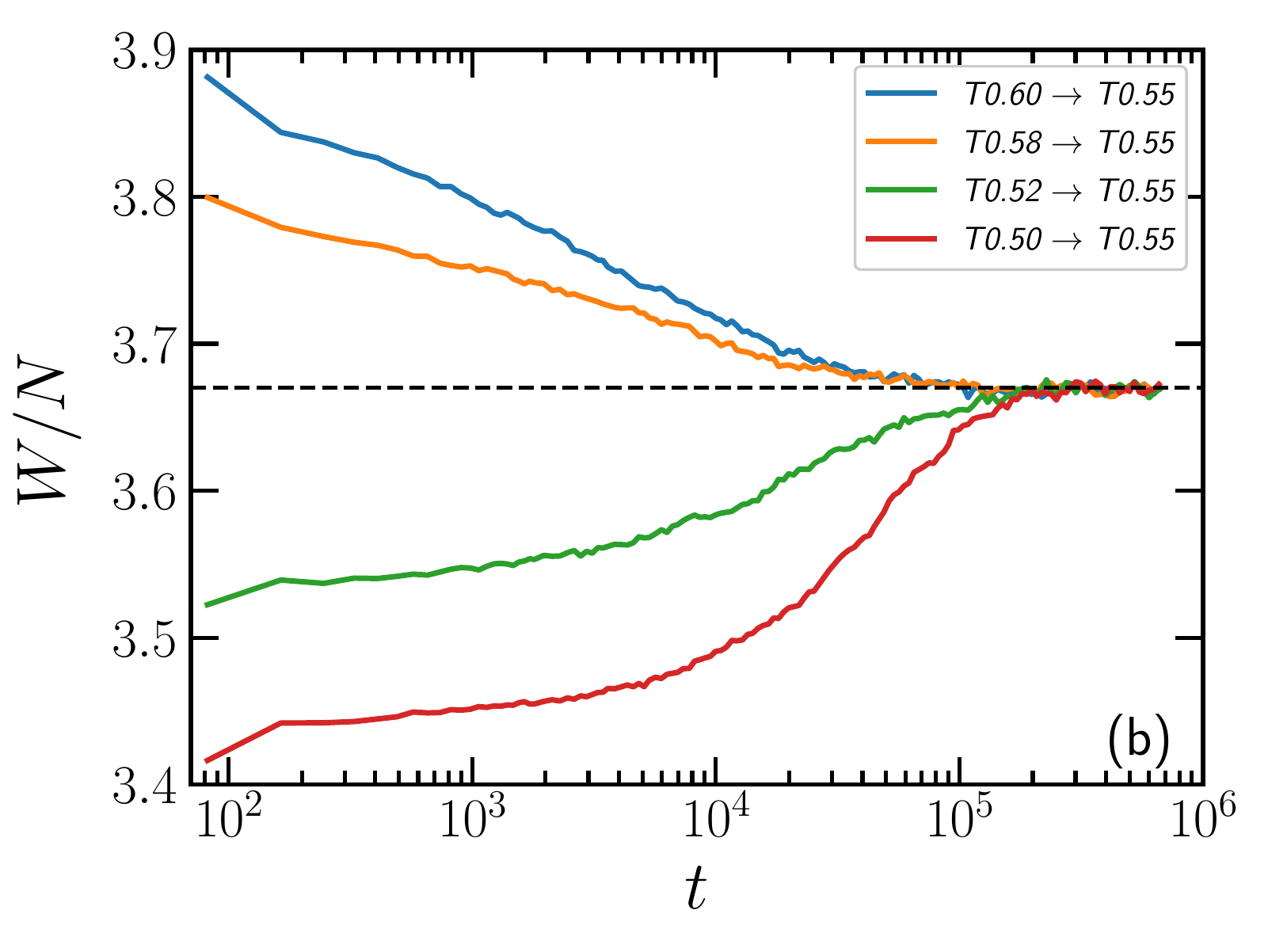}
	\includegraphics[width=7cm]{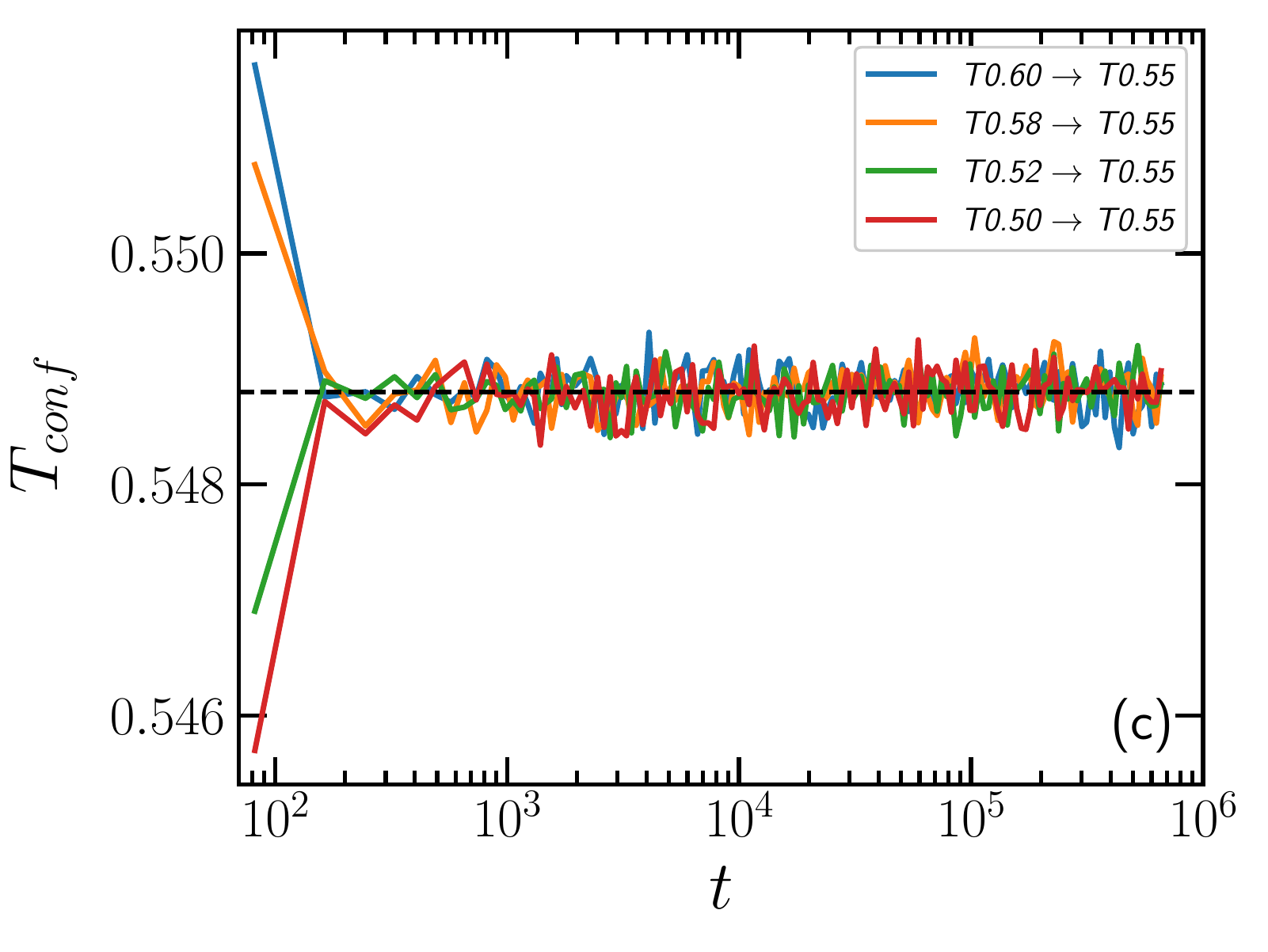}
	\includegraphics[width=7cm]{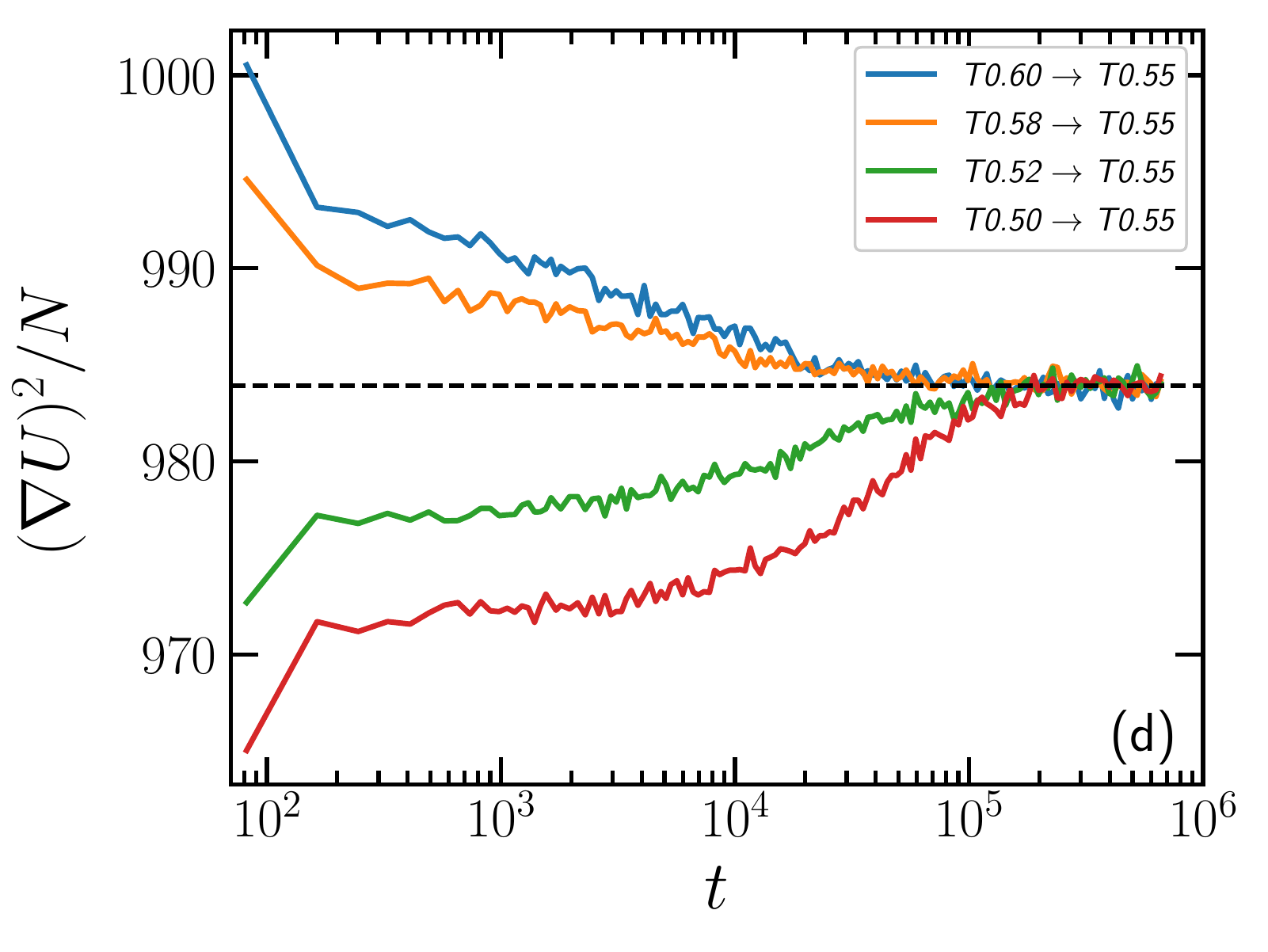}
	\includegraphics[width=7cm]{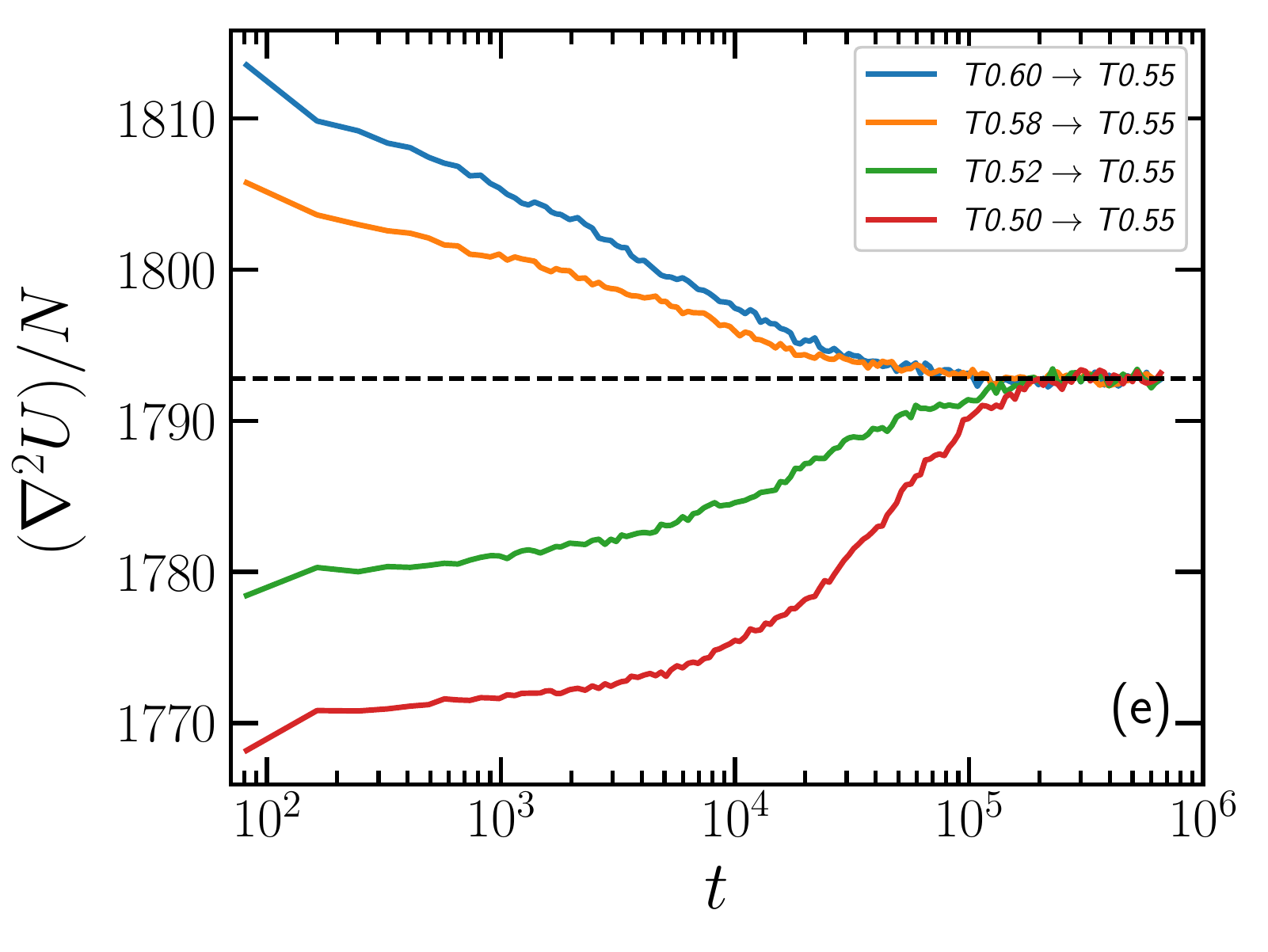}	
	\caption{\label{aging_data}	Aging data for jumps to the same target temperature. Each panel shows four jumps from $T_{\rm start}=T_0+\Delta T$ to $T_0=0.55$ with $\Delta T = \pm 0.03$ and $\Delta T =\pm 0.05$. Results for temperature up jumps are shown in red and green, while down jumps are blue and orange. The fictive temperature effect (asymmetry of approach) is clearly observed, with down jumps being faster and more stretched than up jumps. (a) potential energy; (b) virial; (c) configurational temperature (\eq{eq1a}), which does not age; (d) average squared force; (e) Laplacian of the potential energy.}
\end{figure}

\begin{figure}[htbp!]
	\includegraphics[width=7cm]{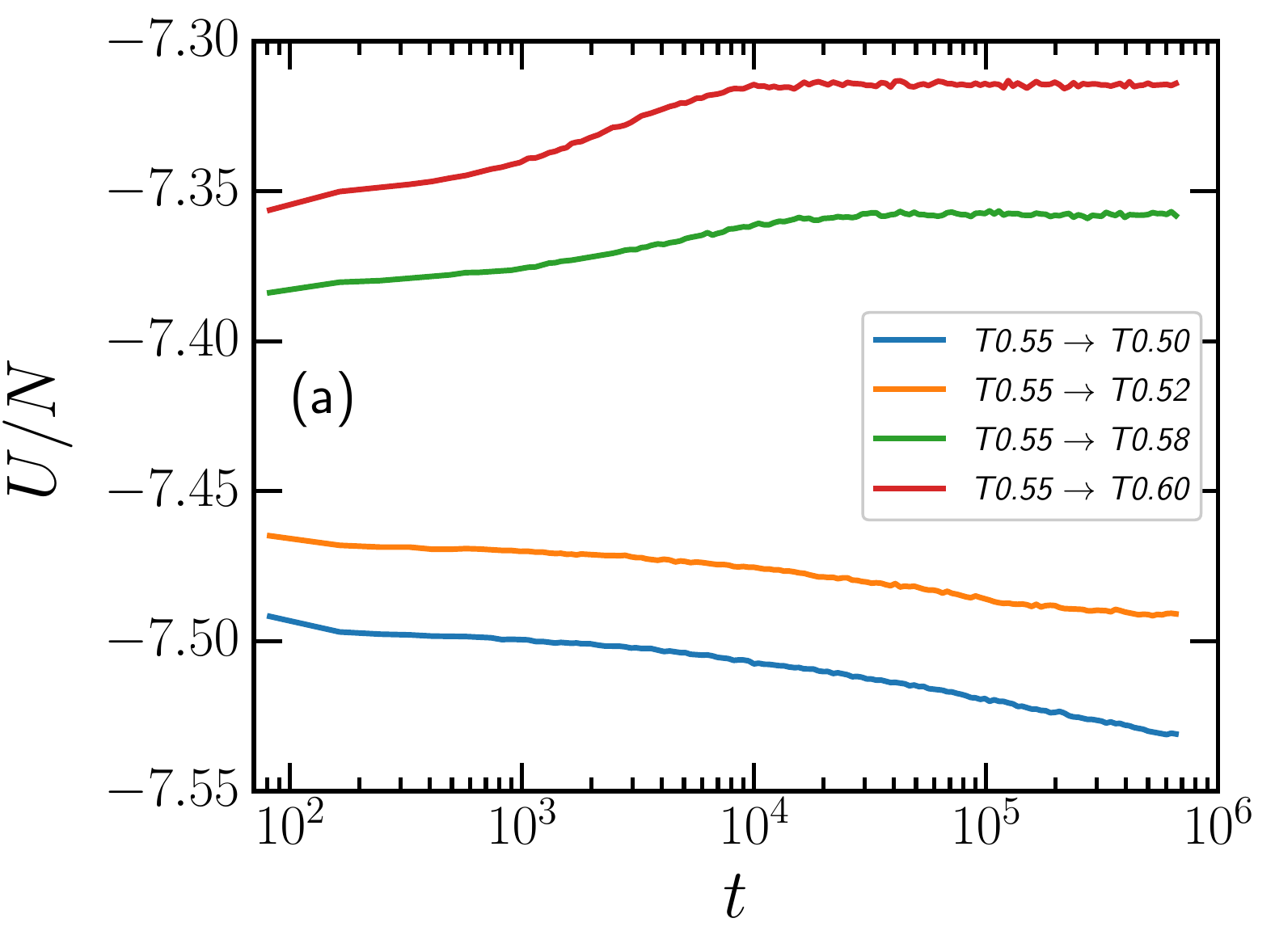}
	\includegraphics[width=7cm]{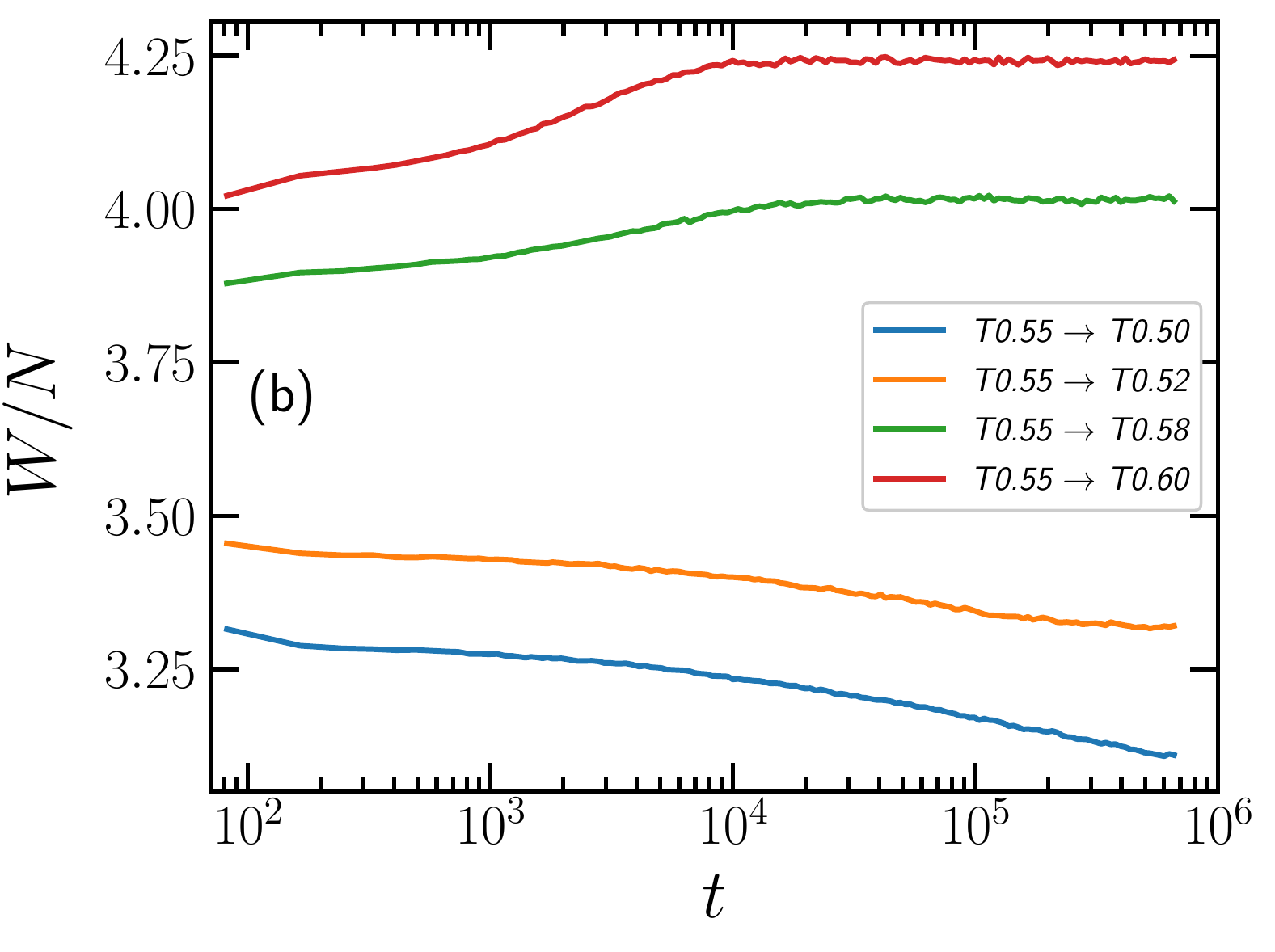}
	\includegraphics[width=7cm]{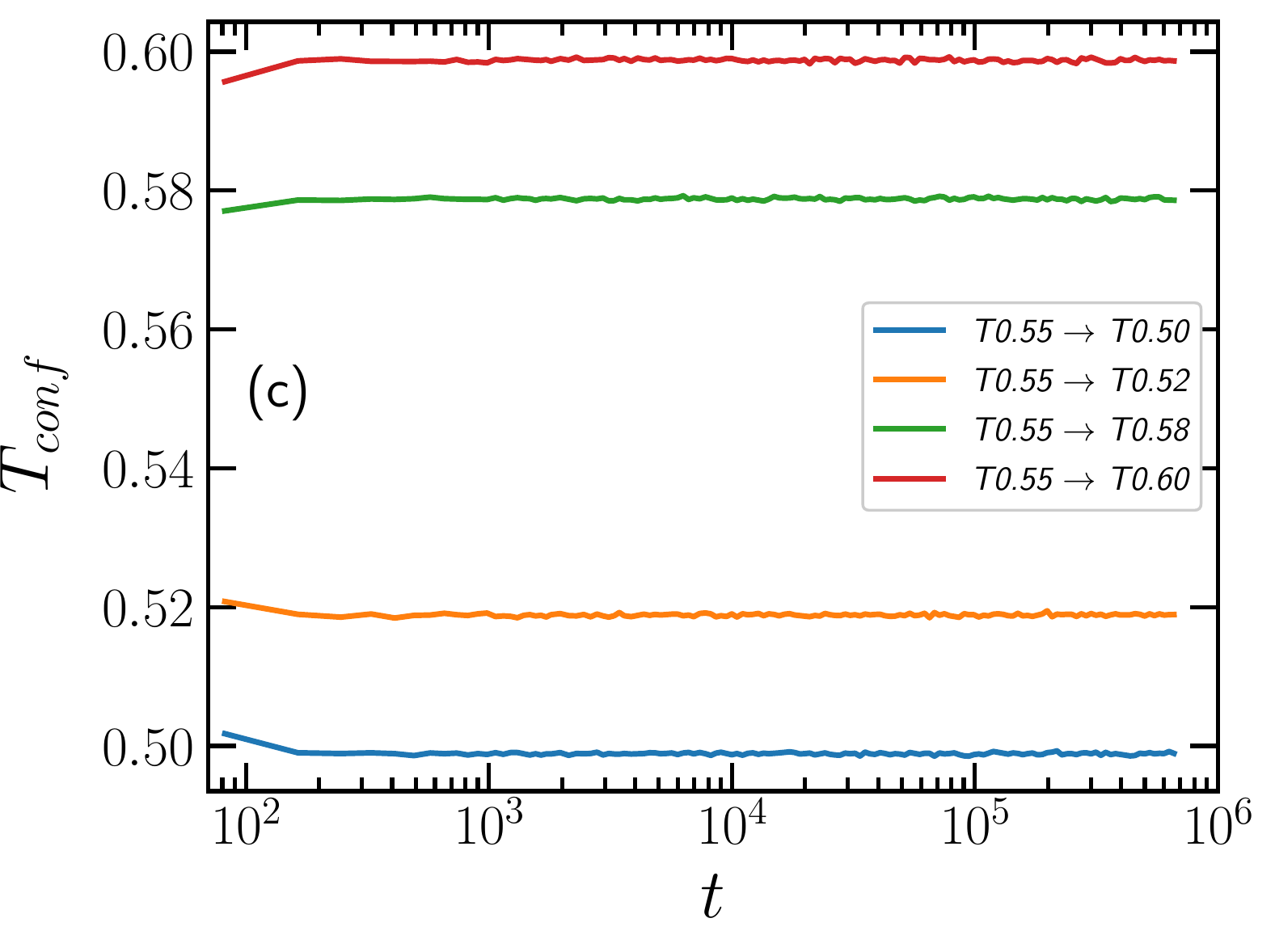}
	\includegraphics[width=7cm]{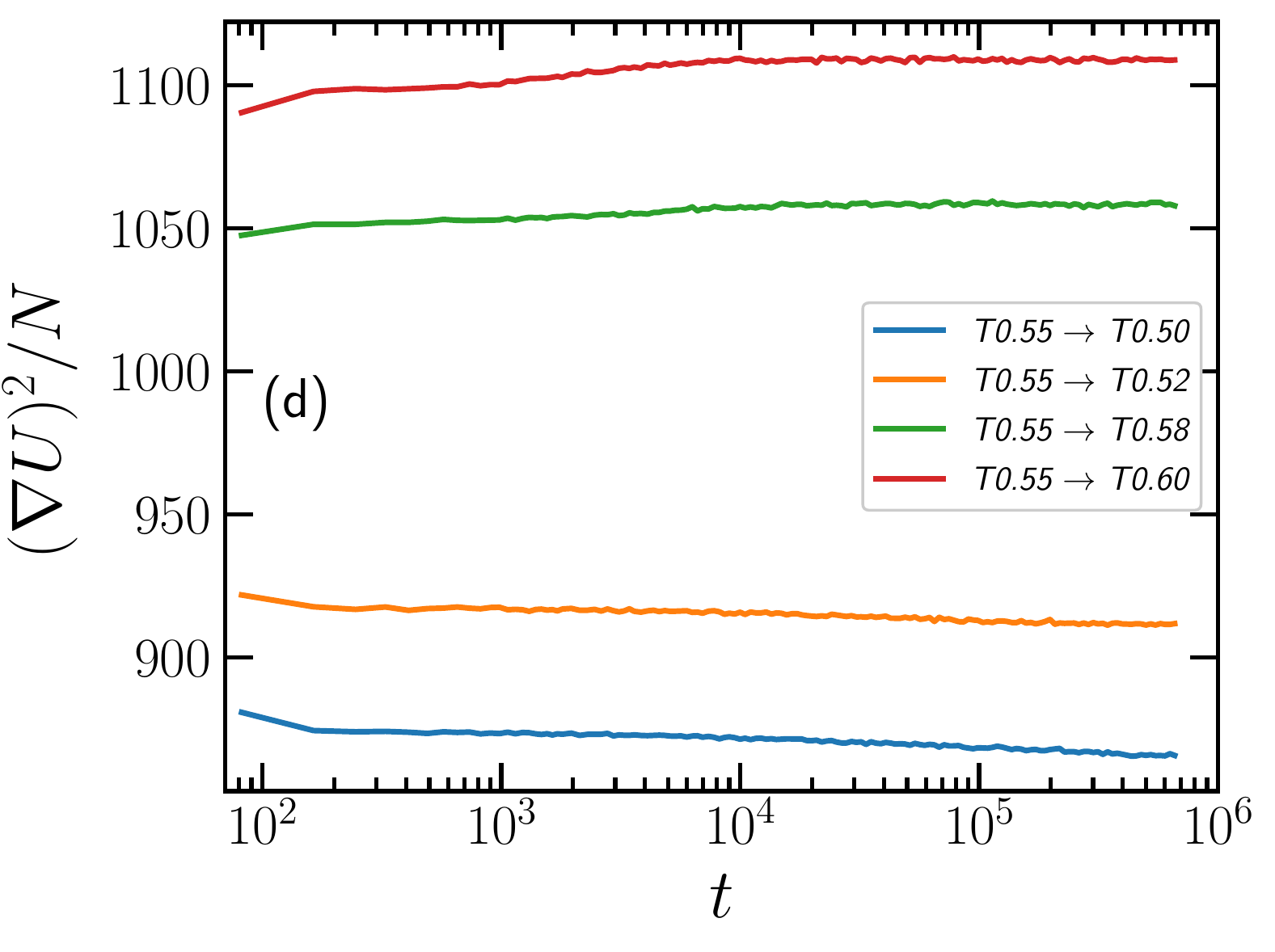}
	\includegraphics[width=7cm]{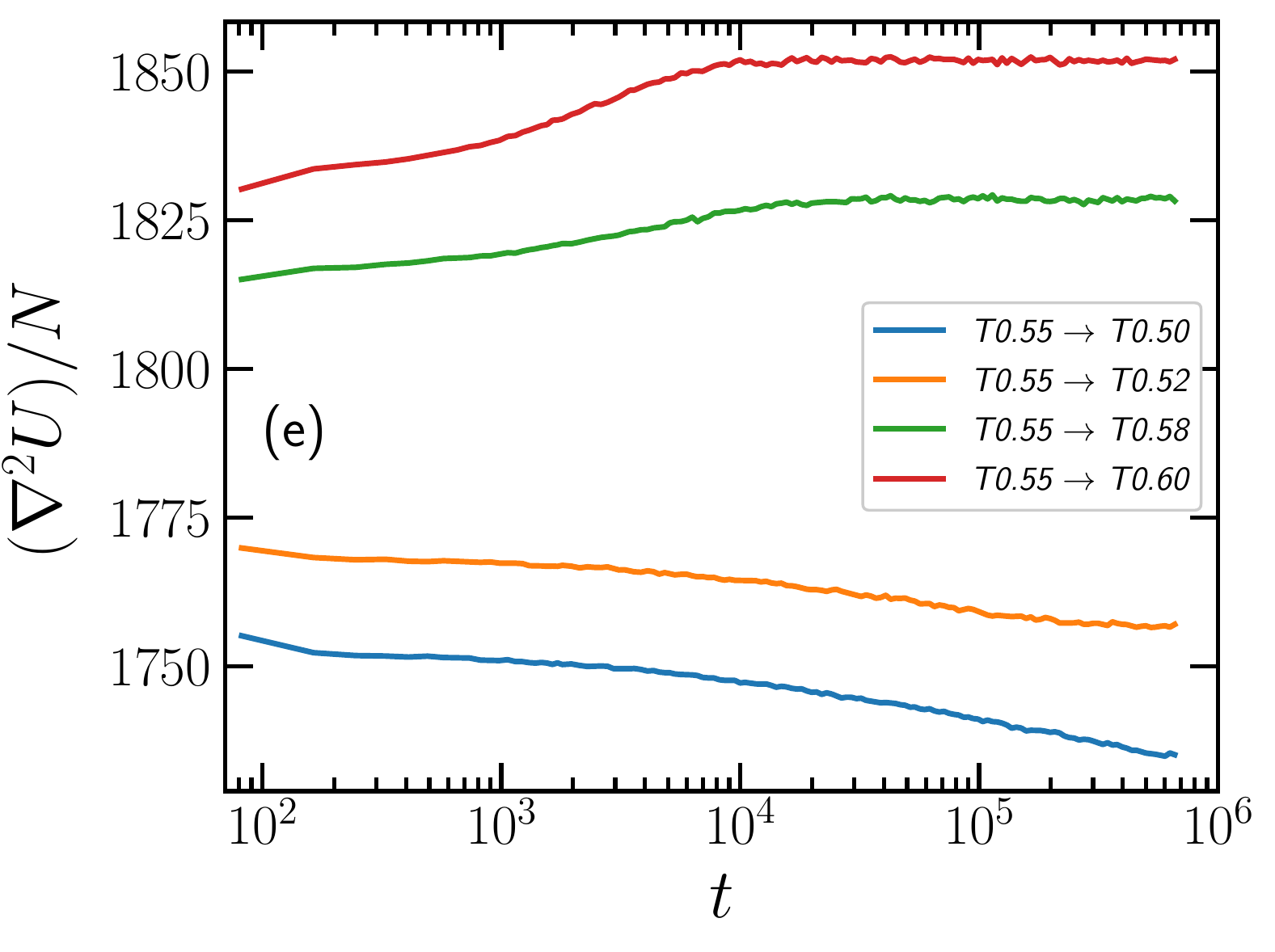}	
	\caption{\label{aging_data_2} Aging data for jumps from the same temperature. Each panel shows four jumps from  $T_0=0.55$ to $0.55+\Delta T$ with $\Delta T = \pm 0.03$ and $\Delta T =\pm 0.05$. Results for temperature up jumps are shown in red and green, while down jumps are blue and orange. Note that the scale on the y-axis is different from that of \fig{aging_data}. (a) potential energy; (b) virial; (c) configurational temperature; (d) average squared force; (e) Laplacian of the potential energy.}
\end{figure}

Initially, the following five quantities were probed: the potential energy, the virial, the configurational temperature defined by

\be\label{eq1a}
k_BT_{\rm conf}\,=\,\frac{\langle (\nabla U)^2 \rangle}{\langle \nabla^2 U \rangle}\,,
\ee
its numerator (the average squared force), and its denominator (the Laplacian of the potential energy). The data of the present study, obtained after averaging over 100 simulations, are presented in \fig{aging_data} and  \fig{aging_data_2}. We find that the configurational temperature does not age but equilibrates almost instantaneously (Fig. 1(c)), confirming previous results by Powles \textit{et al.} \cite{pow05}. For this reason, the remainder of the paper focuses on the aging of the four other quantities. These quantities are easily probed and obvious choices for testing SPA in a computer simulation.

After equilibration at each starting temperature $T_{\rm start}=T_0+\Delta T$, we initiate an aging simulation at $t=0$ by changing the thermostat temperature to the ``target'' temperature $T_0$. The system eventually reaches thermal equilibrium at  $T_0$. We denote the quantity probed by $\chi(t)$. The equilibrium value of $\chi$ at $T_0$ is denoted by $\chi _{\rm eq}$, while $\chi(0)$ is the equilibrium value of $\chi$ at $T_{\rm start}$, i.e., just before the jump is initiated at $t=0$.

From $\chi(t)$ we define for each temperature jump the normalized relaxation function $R(t)$ by subtracting the value of $\chi$ at $T_0$ from the value at each time, subsequently dividing by the overall change, i.e.,

\be\label{R_t}
R(t)\,\equiv\,\frac{\chi (t) - \chi_{eq}}{\chi (0) - \chi_{eq}}\,.
\ee
Note that $R(0)=1$ just before the jump is initiated. Within a few time steps after $t=0$ there is a significant ``instantaneous'' drop in $R(t)$. Aging descriptions conventionally focus only on the subsequent, relaxing part of the temperature response, but it is convenient to use instead the above defined  $R(t)$ because this quantity can be determined directly from the data without a need to estimate the magnitude of the initial ``instantaneous'' change of $\chi$.

\section{Single-parameter aging}

We briefly review here the derivation of SPA, which is based on two assumptions within the TN formalism \cite{Hecksher2015,roed2019generalized}. The first assumption is that the clock rate, $\gamma(t)$, is determined by the monitored parameter $\chi(t)$ itself. The second assumption is that temperature changes are so small that a first-order Taylor expansion of the logarithm of the aging rate in terms of $\chi$ applies. If $\Delta\chi(t)\equiv\chi(t)-\chi_{\rm eq}$ is the variation of $\chi$ from its equilibrium value at the target temperature $T_0$ (implying that $\Delta\chi(t)\to 0$ as $t\to\infty$), the first-order Taylor expansion leads to

\be\label{eq4}
\ln\gamma(t)=\ln\gamma_{\rm eq}+\Delta\chi(t)/\chi_{\rm const}
\ee
in which $\gamma_{\rm eq}$ is the equilibrium relaxation rate at the target temperature $T_0$ and $\chi_{\rm const}$ is a constant of same dimension as $\chi$. This expression summarizes the general SPA framework. In conjunction with the TN basic assumption that physical aging is a linear response in the temperature variation when formulated in terms of the material time, SPA may be applied to any relatively small temperature variation, be it continuous or discontinuous. We henceforth consider the simplest case, that of a (discontinuous) temperature jump.

Since $\Delta\chi(t)=\Delta\chi(0)R(t)$ by the definition of $R(t)$, \eq{eq4} implies \cite{Hecksher2015}

\be\label{eq2}
\gamma(t)\,=\,\gamma_{\rm eq} \exp\left(\dfrac{\Delta\chi(0)}{\chi_{\rm const}}R(t)\right)\,.
\ee
The normalized relaxation function $R(t)$ is given by \cite{Narayana1971,scherer1986relaxation,Hecksher2015}

\be\label{eq6a}
R(t)=\Phi(\xi)\,.
\ee
The point of the TN formalism is now that the function $\Phi(\xi)$ is the same for all temperature jumps. In contrast, the time-dependent material time $\xi(t)$ is not at all universal because the aging rate changes as the system ages. In conjunction with the definition of the aging rate in terms of the material time \eq{xi}, \eq{eq6a} implies $\dot{R}(t)=\Phi^\prime(\xi)\gamma(t)$. Since \eq{eq6a} means that $\xi$ is the same function of $R$ for all jumps, by defining $F(R)\equiv - \Phi^\prime(\xi(R))$ one gets \cite{Hecksher2015} 

\be\label{eq3}
\dot{R}(t) 
\,=\,-F(R)\gamma(t)\,.
\ee
The negative sign in Eq. (\ref{eq3}) is convenient because $R(t)$ is (usually) a monotonically decreasing function of time, thus making $F(R)$ positive. 

Equations (\ref{eq2}) and (\ref{eq3}) lead to

\be\label{eq4a}
-\frac{\dot{R}(t)}{\gamma_{eq}} \exp \left(-\dfrac{\Delta \chi(0)}{\chi_{const}}R(t)\right)=F(R)\,.
\ee	
Since the right-hand side is independent of the jump sign and magnitude, this applies also for the left-hand side. This prediction was validated in 2015 in experiments monitoring four different quantities \cite{Hecksher2015}. From \eq{eq4a} one can basically predict the relaxation function for one jump from the relaxation function of another jump since a single jump is enough to determine the function $F(R)$. In order to determine the constant $\chi_{\rm const}$, however, two jumps are needed (see below); alternatively, a determination of the equilibrium relaxation rate at two different temperatures can also be used to find $\chi_{\rm const}$. We refer below to the ``known'' relaxation function as ``jump1'' while the relaxation function to be compared to the prediction based on jump1, is referred to as ``jump2''.

For the times $t_{1}^{*}(R)$ and $t_{2}^{*}(R)$ at which two jumps have the same normalized relaxation function, i.e., $R_1=R_2$, since $F(R_1)=F(R_2)$ Eq. (\ref{eq4a}) implies that 

\be\label{eq5}
-\frac{dR_1}{dt_{1}^{*}}.\dfrac{1}{\gamma_{eq,1}}.\exp \left(-\frac{\Delta\chi(0)_1}{\chi_{const}}R(t_{1}^{*})\right)= -\frac{dR_2}{dt_{2}^{*}}.\dfrac{1}{\gamma_{eq,2}}.\exp \left(-\frac{\Delta\chi(0)_2}{\chi_{const}}R(t_{2}^{*})\right)\,.
\ee
If we choose $dt_{1}^{*}$ and $dt_{2}^{*}$ such that $dR_1=dR_2$ and use $R_1(t_{1}^{*})=R_2(t_{2}^{*})$, \eq{eq5} leads to

\be\label{eq6}
dt_{2}^{*}=\dfrac{\gamma_{eq,1}}{\gamma_{eq,2}}\exp\left(\dfrac{\Delta \chi(0)_1 - \Delta \chi(0)_2 }
{\chi_{const}}R(t_{1}^{*})\right)dt_{1}^{*},
\ee
By integrating this one gets

\be\label{eq7}
t_2=\int_{0}^{t_2}dt_{2}^{*}=\dfrac{\gamma_{eq,1}}{\gamma_{eq,2}}\int_{0}^{t_{1}}
\exp\left(\dfrac{\Delta \chi(0)_1 - \Delta \chi(0)_2 }{\chi_{const}}R(t_{1}^{*})\right)dt_{1}^{*}\,.
\ee
Equation (\ref{eq7}) states that for predicting jump2 one just needs to ``transport'' the discrete time vector $\mathbf{t}_1=(t_1^1, t_1^2,...,t_1^n)$ and its corresponding relaxation vector $\mathbf{R}_1=(R_1^1,R_1^2,...,R_1^n)$ to a new time vector $\mathbf{t}_2=(t_2^1, t_2^2,...,t_2^n)$, corresponding to the same $\mathbf{R}$ vector $\mathbf{R}_1$ \cite{roed2019generalized}. Thus by plotting $(\mathbf{t}_2,\mathbf{R}_1)$ and $(\mathbf{t}_2,\mathbf{R}_2)$, data are predicted to collapse if SPA applies. For jumps to the same target temperature Eq. (\ref{eq7}) reduces to \cite{Hecksher2015}

\be\label{eq8}
t_2=\int_{0}^{t_{1}}\exp\left(\dfrac{\Delta \chi(0)_1 - \Delta \chi(0)_2}{\chi_{const}}R(t_{1}^{*})\right)dt_{1}^{*}\,.
\ee

\begin{figure}[h]
	\includegraphics[width=7cm]{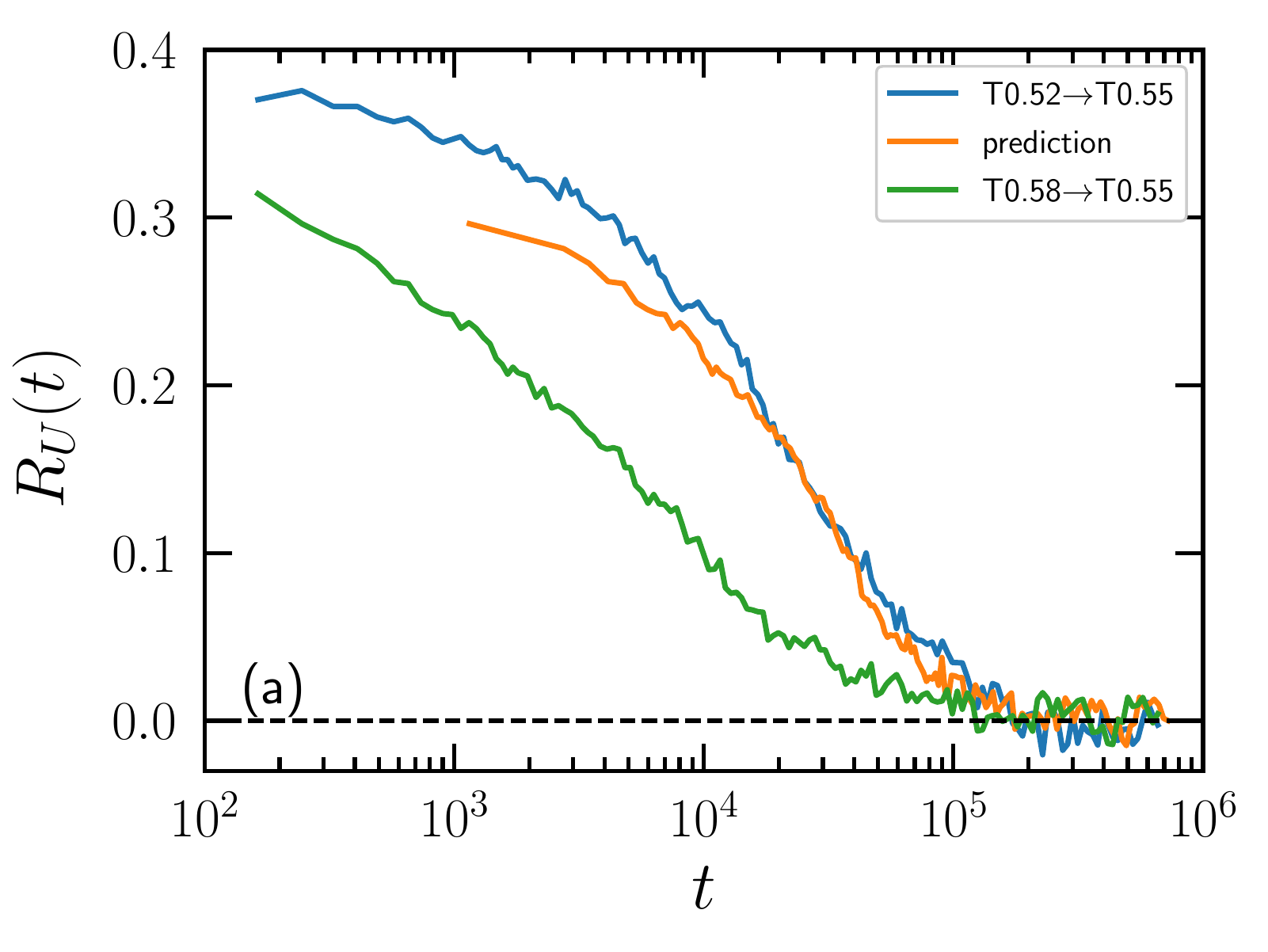}
	\includegraphics[width=7cm]{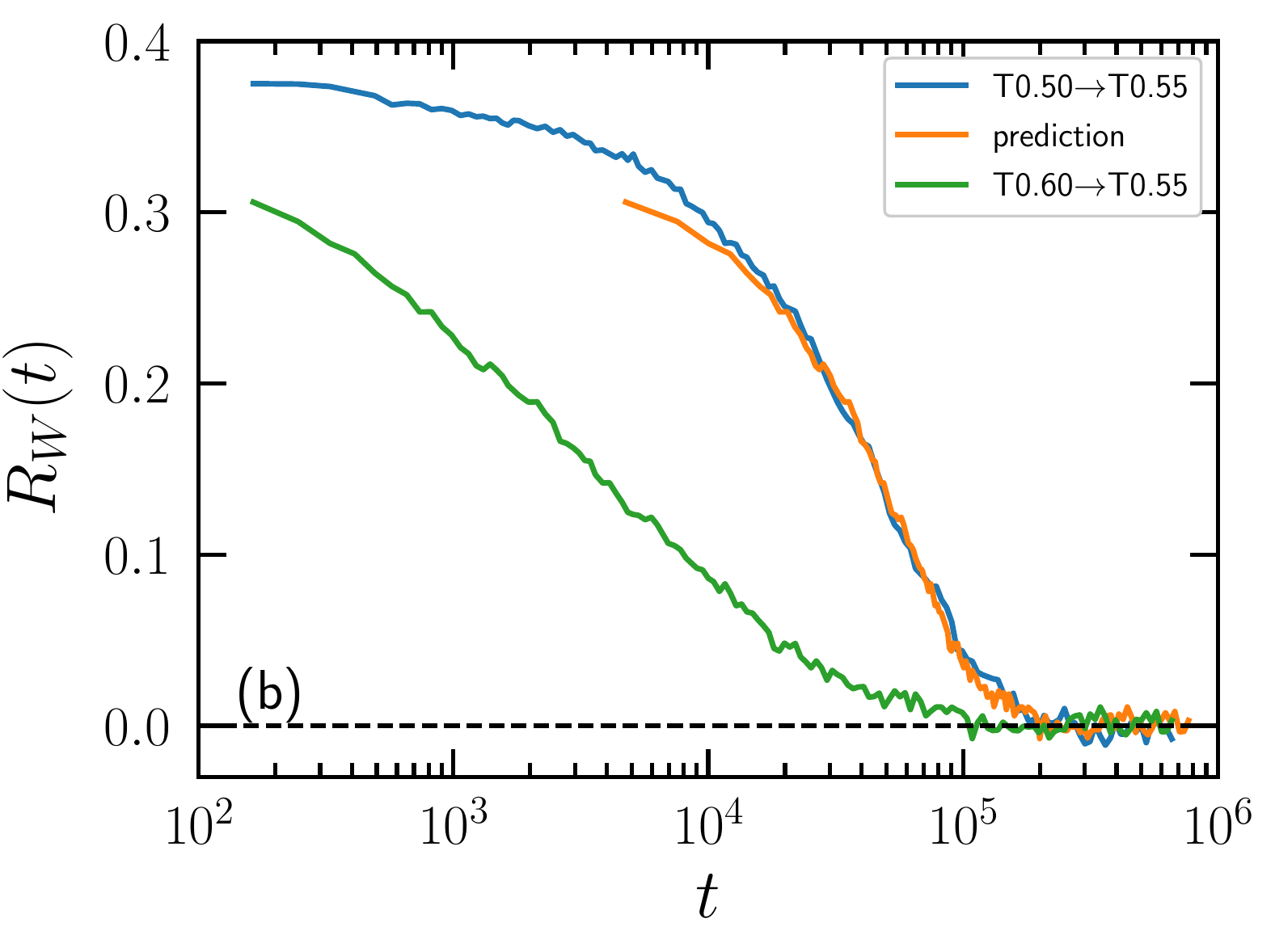}
	\includegraphics[width=7cm]{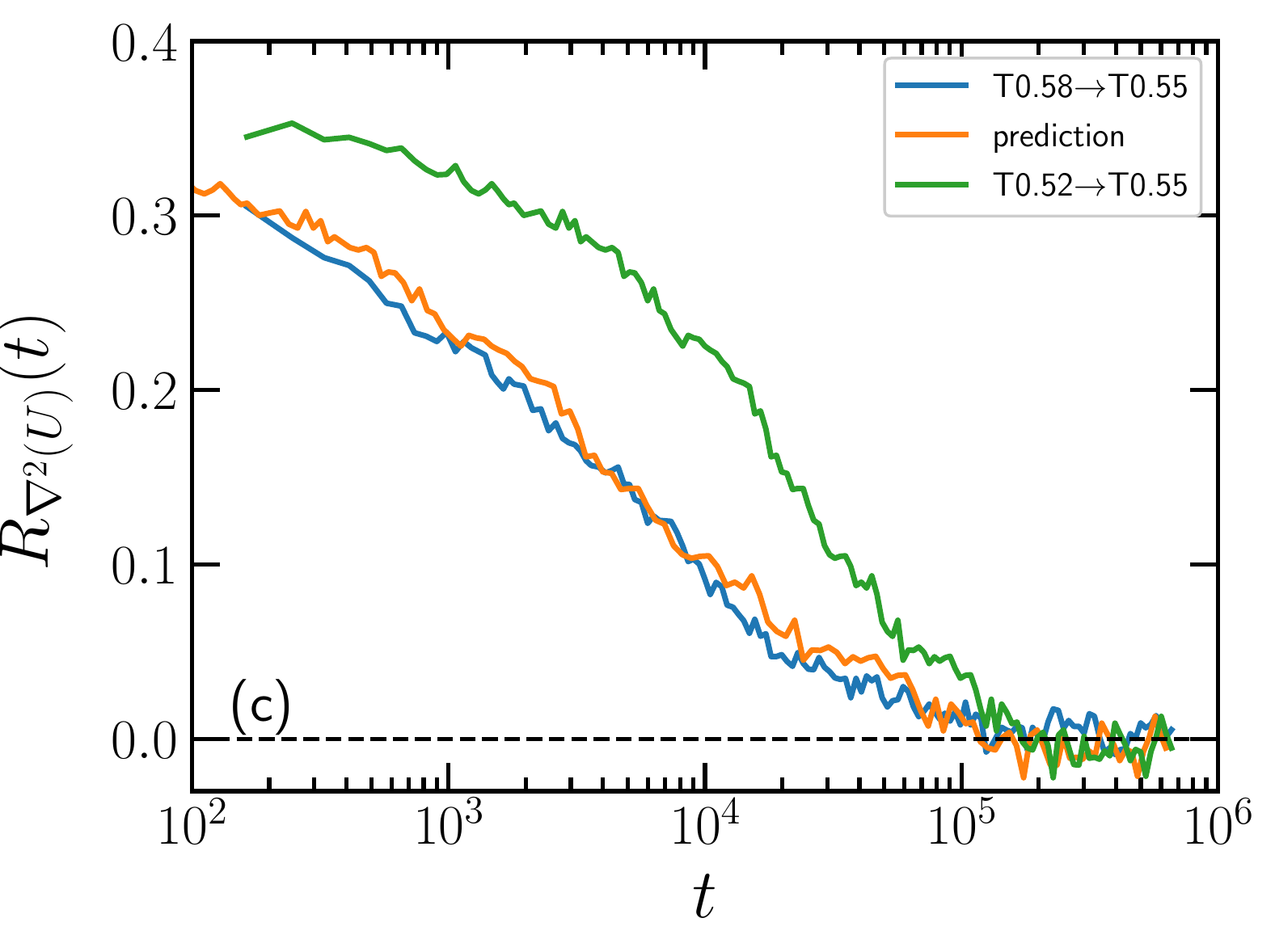}
	\includegraphics[width=7cm]{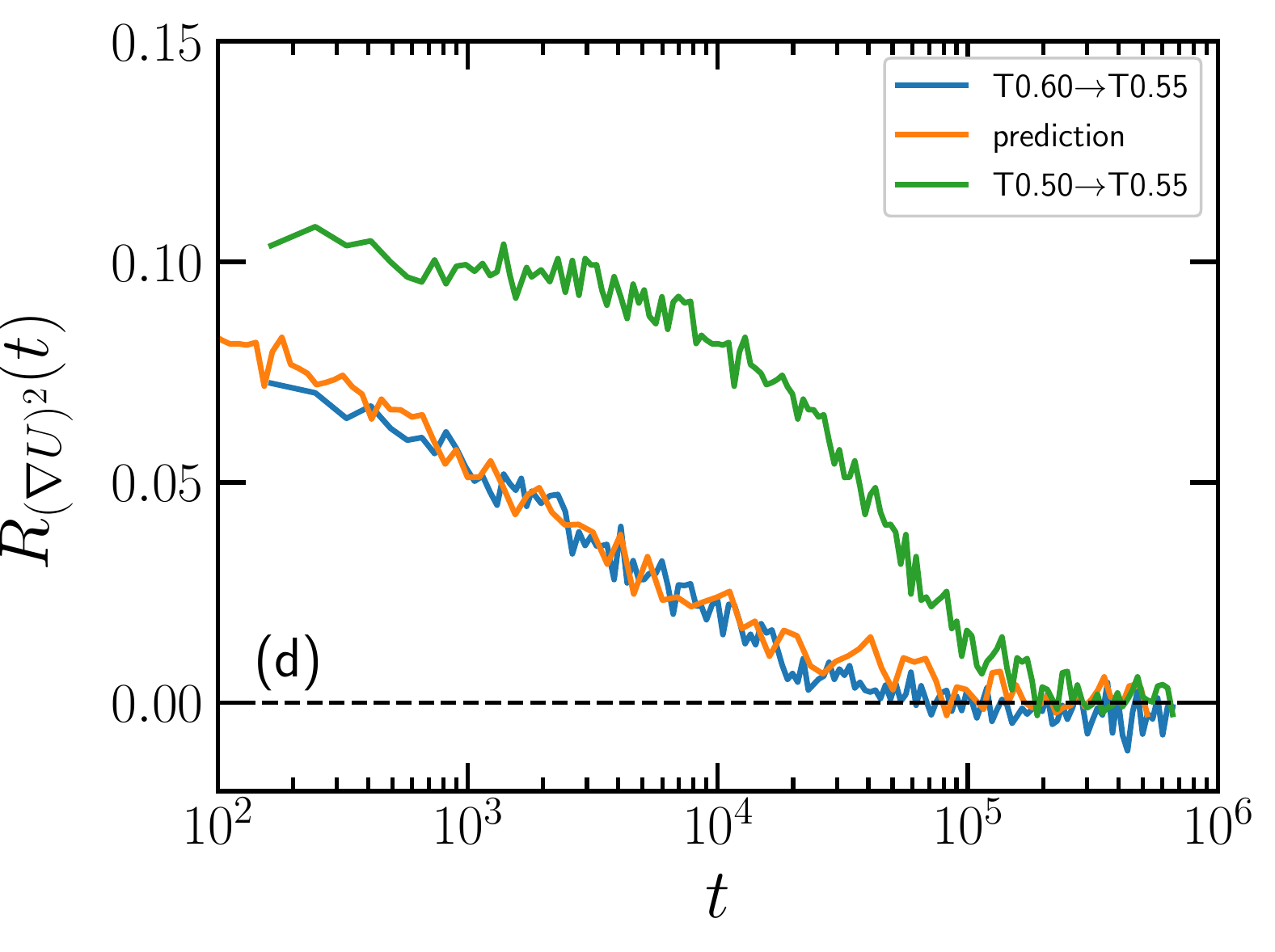}
	\caption{\label{prediction1} Test of the SPA predictions for jumps to the same target temperature $T_0=0.55$. The data for the ``jump1'' normalized relaxation functions $R(t)$ are given by the green curves. The predictions based on jump1 according to Eq. (\ref{eq8}) are given by the orange curves. These are to be compared to the ``jump2'' data (blue curves). (a) and (b) give the predictions of up jumps based on down jumps for the potential energy and the virial, respectively. (c) and (d) give the predictions of down jumps based on up jumps for the Laplacian of the potential energy and the average squared force, respectively. }
\end{figure}

\begin{figure}[h]
	\centering
	\includegraphics[width=7cm]{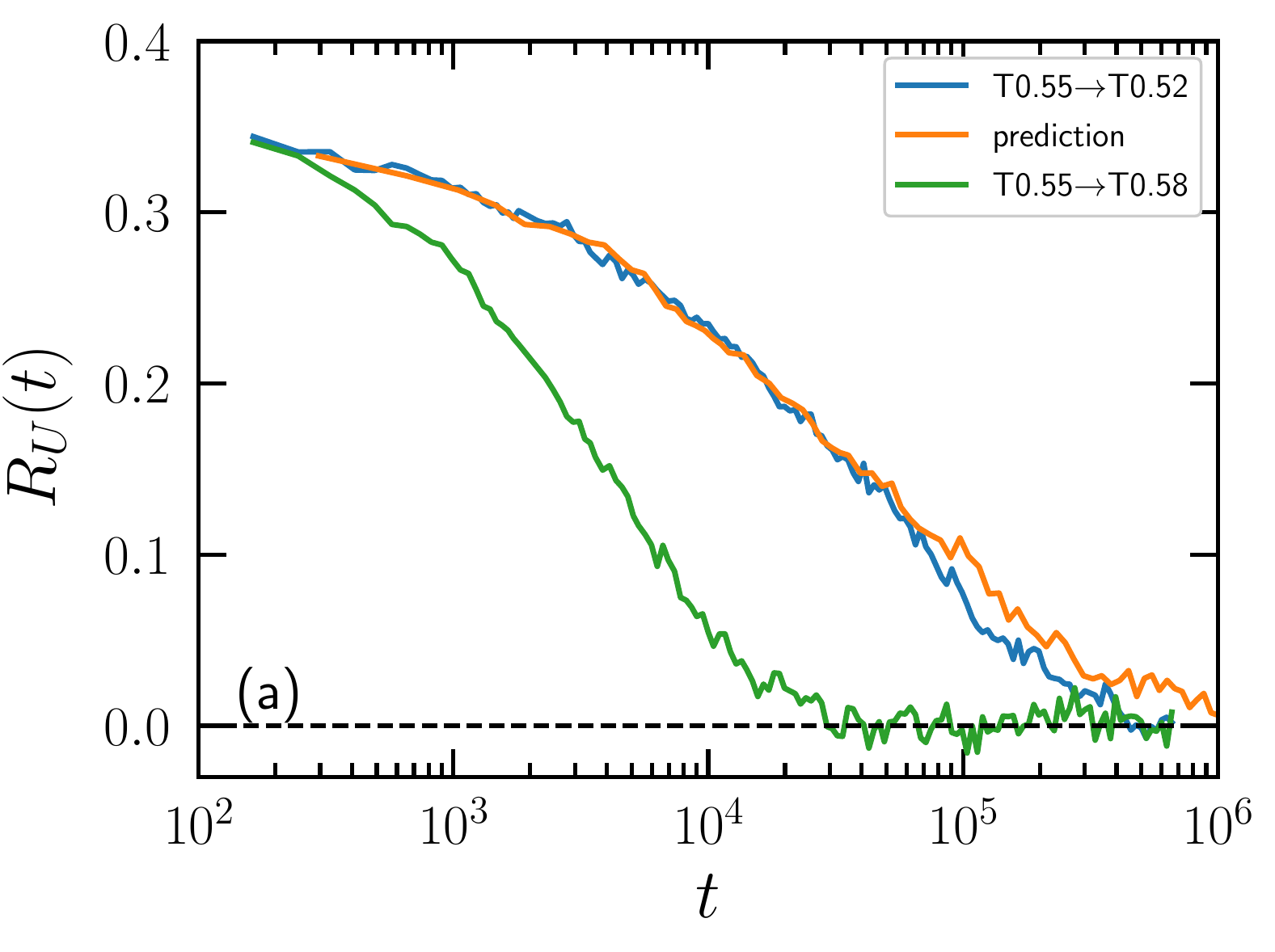}
	\includegraphics[width=7cm]{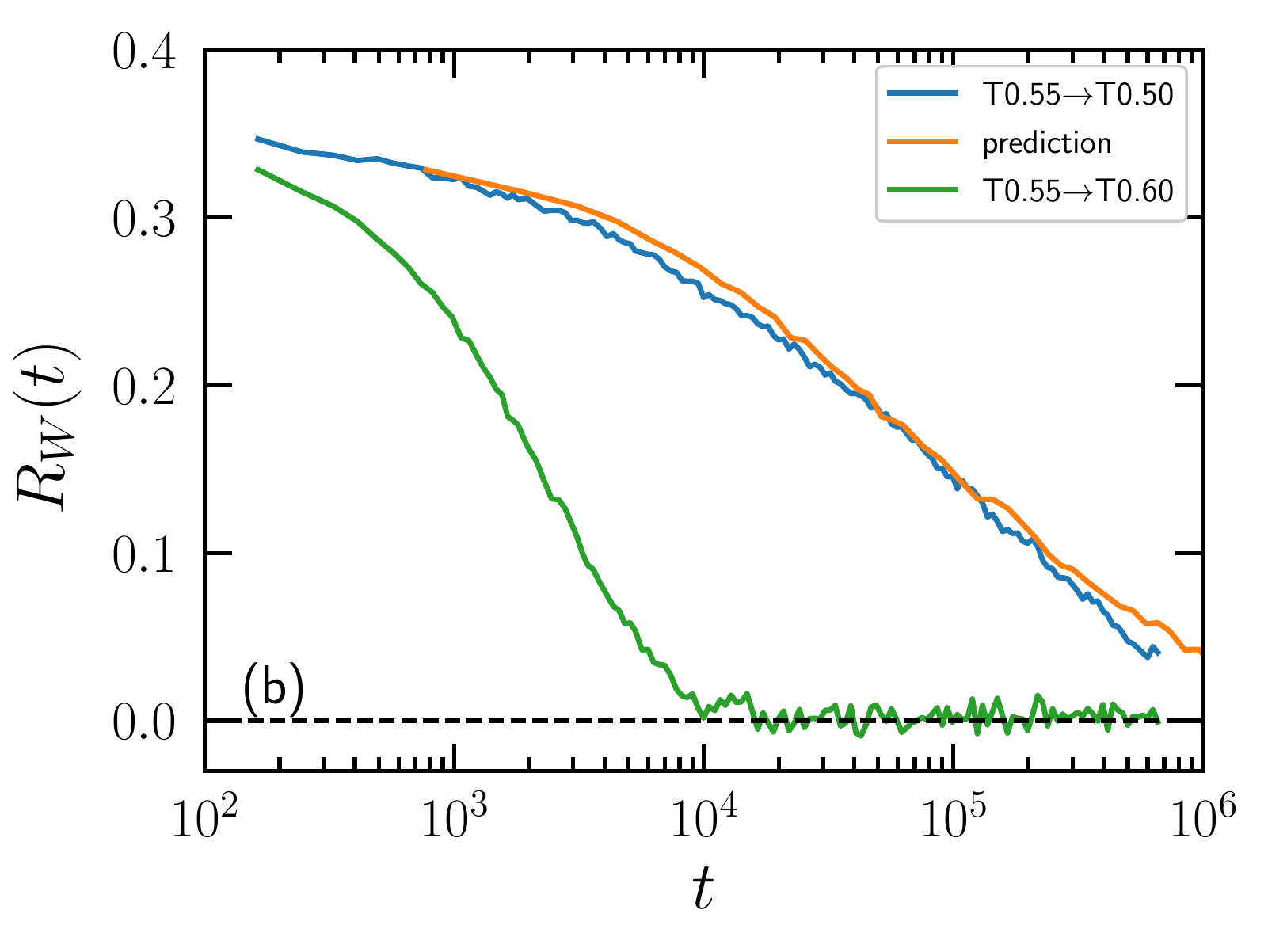}
	\includegraphics[width=7cm]{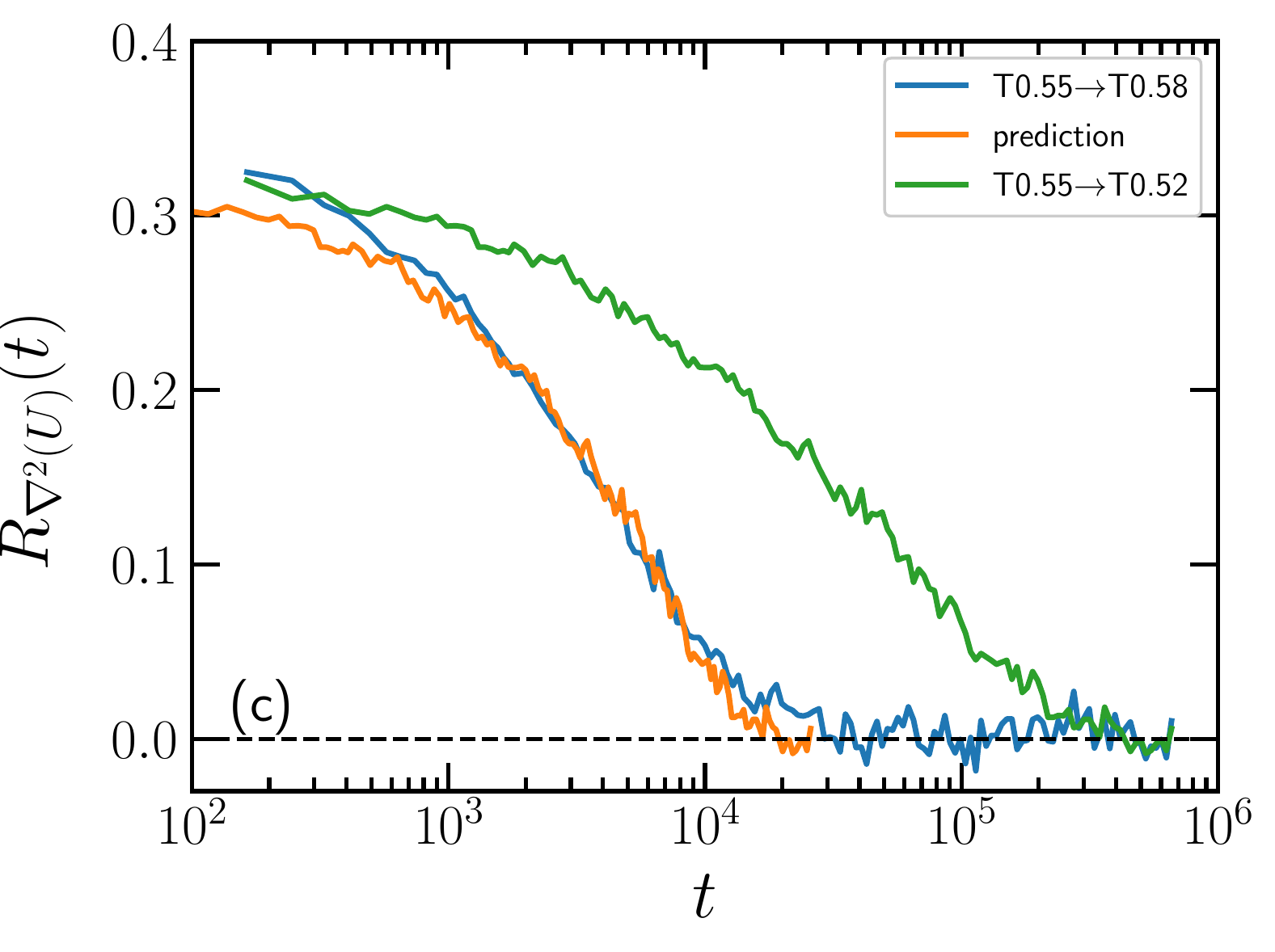}
	\includegraphics[width=7cm]{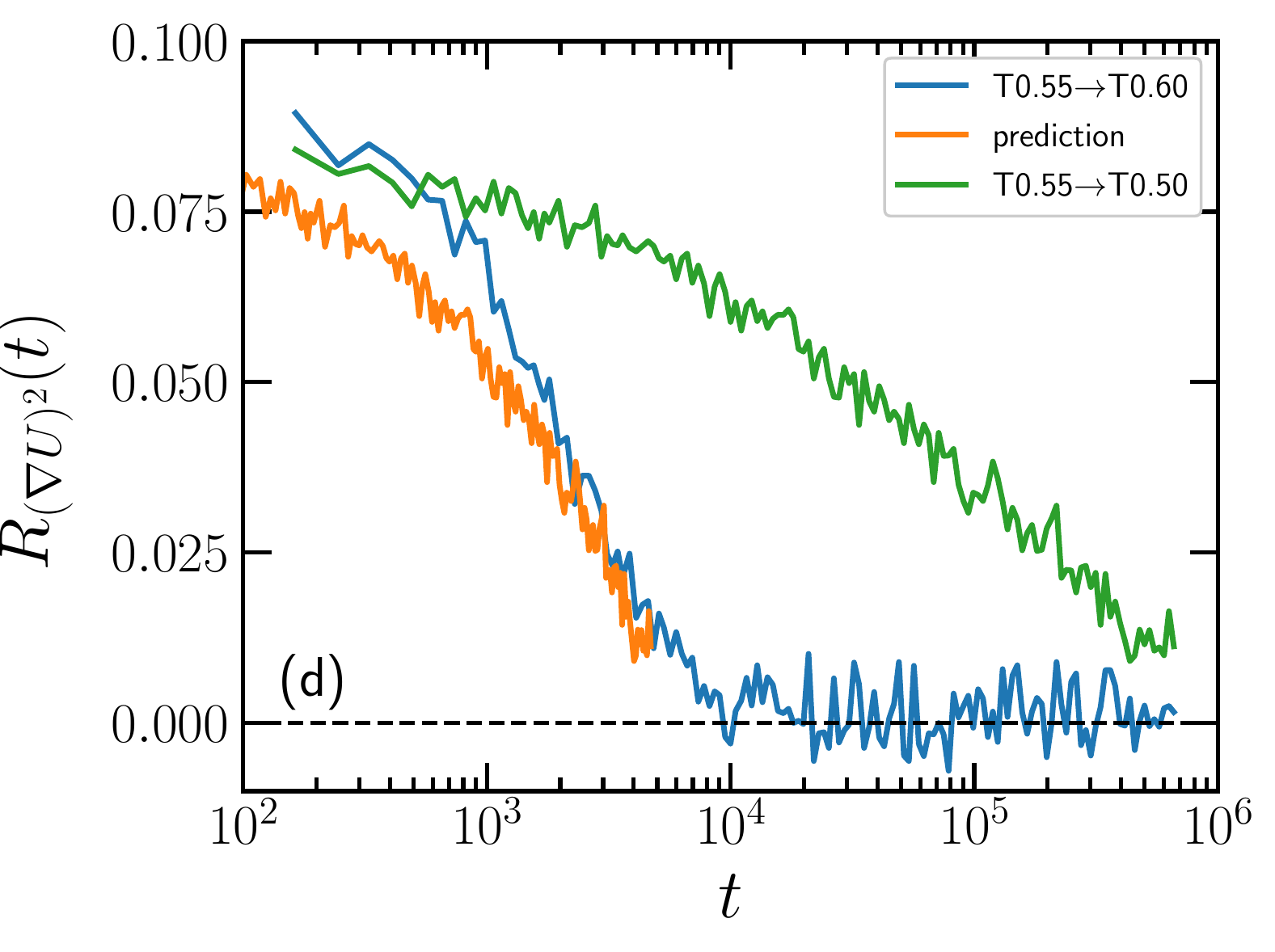}
	\caption{\label{prediction2} SPA tested for jumps starting at the same temperature ($0.55$) and ending at different target temperatures. Jump1 data are the green curves. The predictions based on jump1 according to Eq. (\ref{eq8}) are orange while the jump2 simulation results are blue. (a) and (b) give the predictions of down jumps based on up jumps for the potential energy and the virial, respectively. (c) and (d) give the predictions for up jumps based on down jumps for the Laplacian of the potential energy and the average squared force, respectively.}
\end{figure}

The more general SPA version developed by Roed \textit{et al.} \cite{roed2019generalized} allows one to predict all jumps from the knowledge of a single jump and $\chi_{const}$ (still assuming that $\Delta T$ is small enough to justify the first-order Taylor expansion \eq{eq4}). In contrast to the first SPA derivation considering only jumps to the same target temperature \cite{Hecksher2015}, however, one needs to know the equilibrium clock rate, $\gamma_{eq}$, at the target temperature $T_0$. In the present paper we identified this quantity from $\gamma_{eq}\equiv{1}/{\tau}$, in which the relaxation time $\tau$ is determined from the intermediate scattering function evaluated at the wave vector corresponding to the first-peak maximum of the AA particle radial distribution function ($\tau$ is taken as the time at which this quantity has decayed to $0.2$).

From two jumps to the same target temperature, by means of Eq. (\ref{eq8}) $\chi_{\rm const}$ can be determined and subsequently used to predict all the other jumps. Thus, Eq (\ref{eq8}) implies that

\be\label{t21}
t_2(R)-t_1(R) = \int_{0}^{t_1(R)} \left[\exp\left(\dfrac{\Delta \chi(0)_1 - \Delta \chi(0)_2 }{\chi_{const}}R(t_{1}^{*}) \right)-1\right]dt_1^*\,.
\ee
A similar expression applies for $t_1(R)-t_2(R)$. Taking the long-time limits of these expressions leads to the self-consistency requirement \cite{Hecksher2015}

\be\label{param}
\int_{0}^{\infty} \left[\exp\left(\dfrac{\Delta \chi(0)_1 - \Delta \chi(0)_2 }{\chi_{\rm const}}R(t_{1}^{*}) \right)-1\right]dt_1^* 
+ \int_{0}^{\infty} \left[\exp\left(\dfrac{\Delta \chi(0)_2 - \Delta \chi(0)_1 }{\chi_{\rm const}}R(t_{1}^{*}) \right)-1\right]dt_2^* = 0\,.
\ee
Equation (\ref{param}) is an equation for $\chi_{const}$ that is easily solved numerically. The value of $\chi_{\rm const}$ depends on the quantity probed, of course. Table \ref{tbl1} provides the values of the $\chi_{\rm const}$ for the four different quantities monitored.

\begin{table}
	\centering
	\caption{Different values of $\chi_{const}$ -- derived using \eq{param} for jumps from 0.60 and 0.50 to the target temperature 0.55.}
	\begin{tabular}{c||c|c|c|c}
		\hline \hline
		quantity & $U$ & $W$ &  $\left(\nabla U\right)^2$  & $\nabla ^2 U$ \\
		\hline
		$\chi_{const}$ &  0.018569 & 0.09944 & 5.117 & 10.04 \\
		\hline
	\end{tabular}
	\vspace{1ex}
	\label{tbl1}
\end{table}

\begin{figure}[htbp!]
	\centering
	\includegraphics[width=7cm]{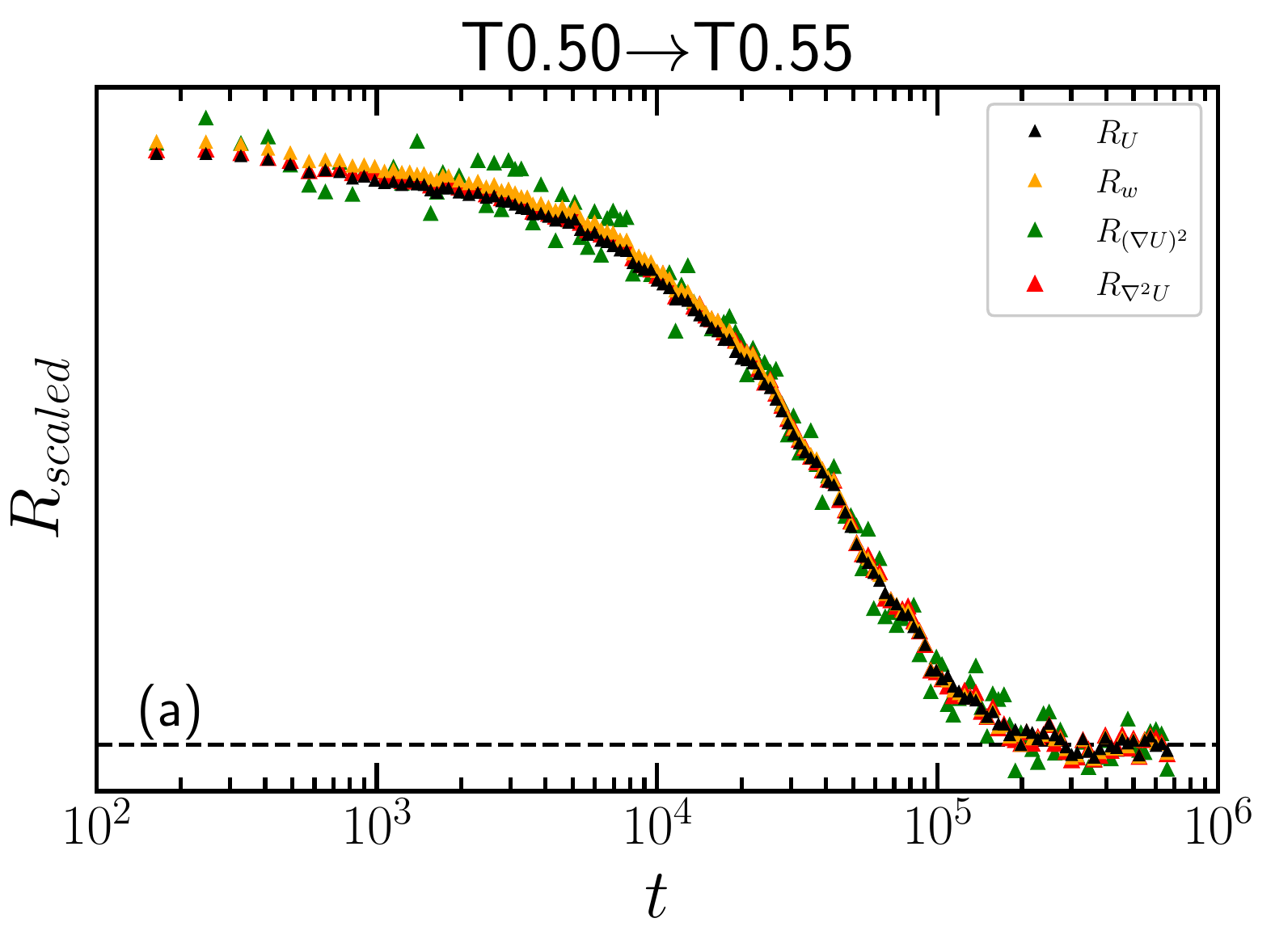}
	\includegraphics[width=7cm]{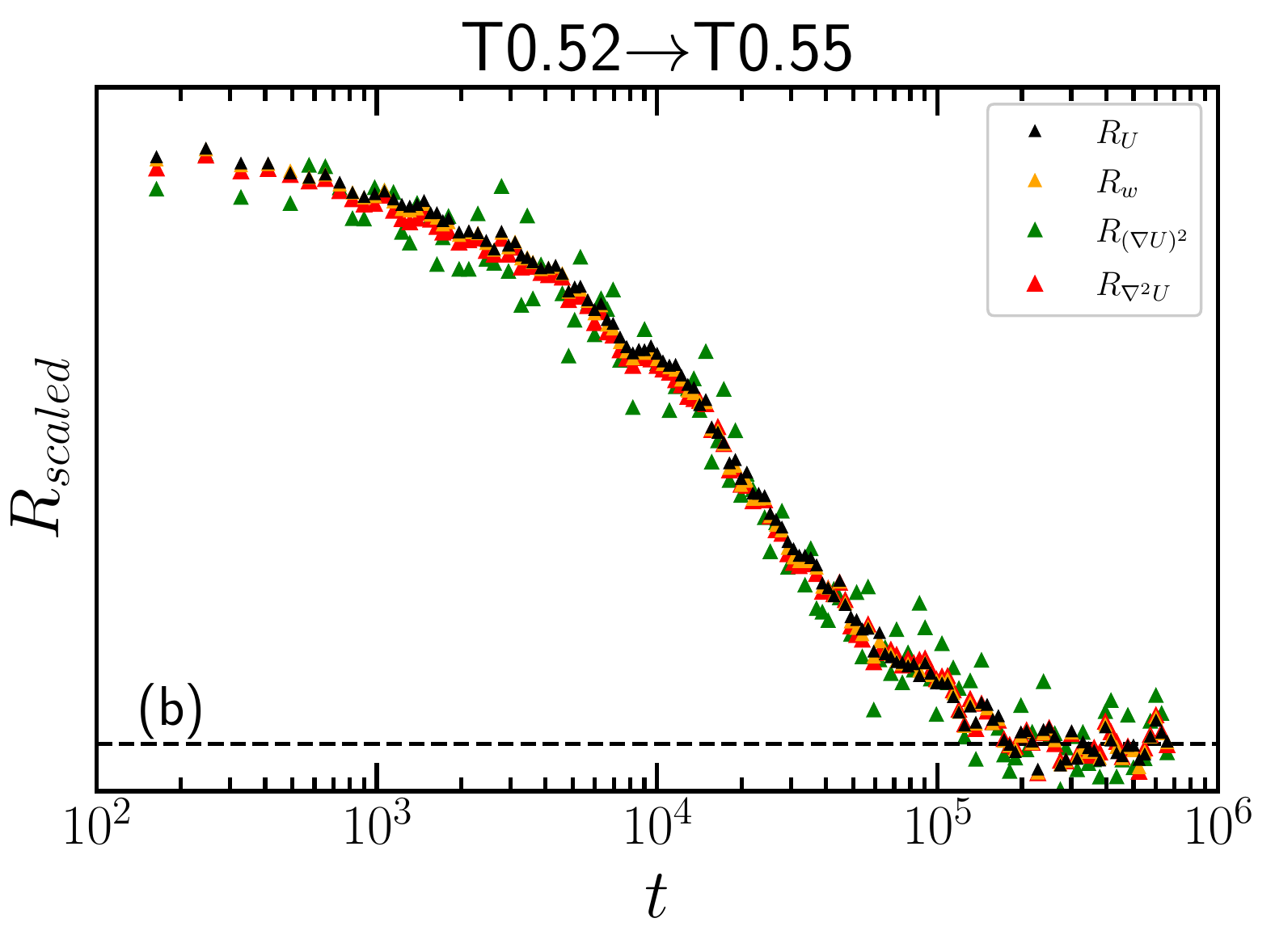}
	\includegraphics[width=7cm]{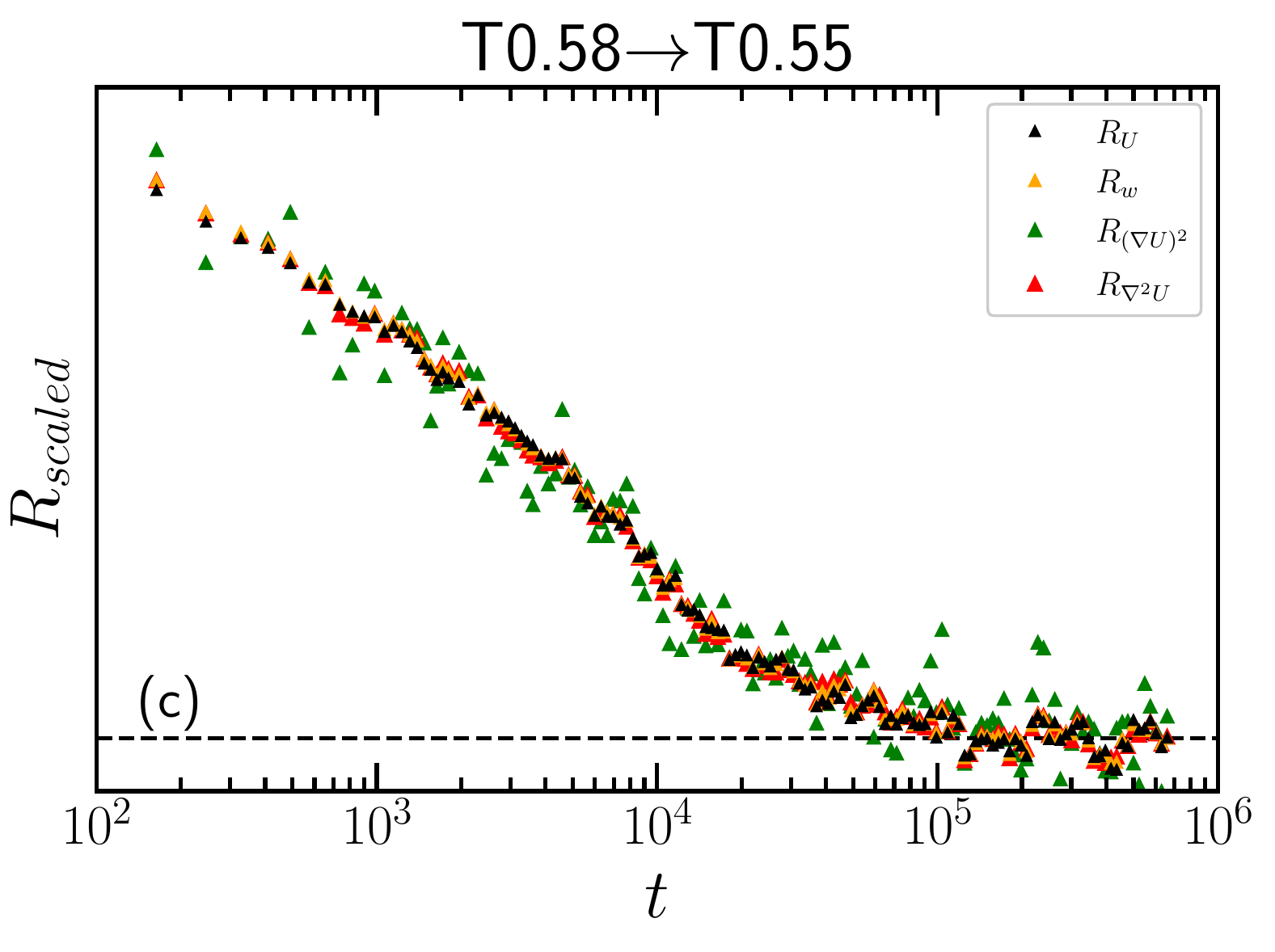}
	\includegraphics[width=7cm]{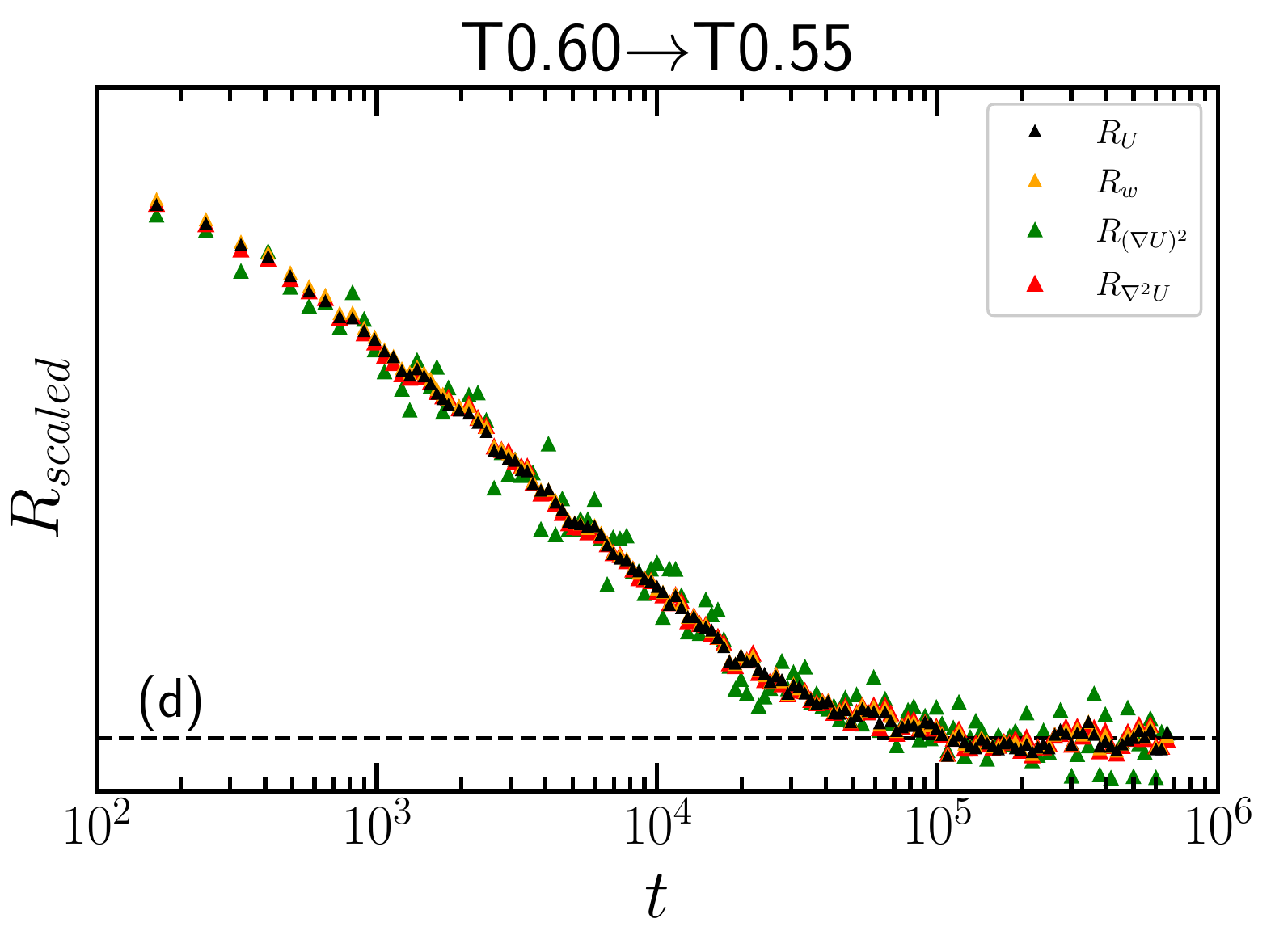}
	\caption{\label{relaxations_upjumps}Empirically scaled relaxation curves of the four quantities, plotted for each of the four jumps to the target temperature $0.55$. The black points represent the potential energy, the virial is yellow, the average squared force is green, and the Laplacian of the potential energy is red.}
\end{figure}

\begin{figure}[htbp!]
	\centering	
	\includegraphics[width=7cm]{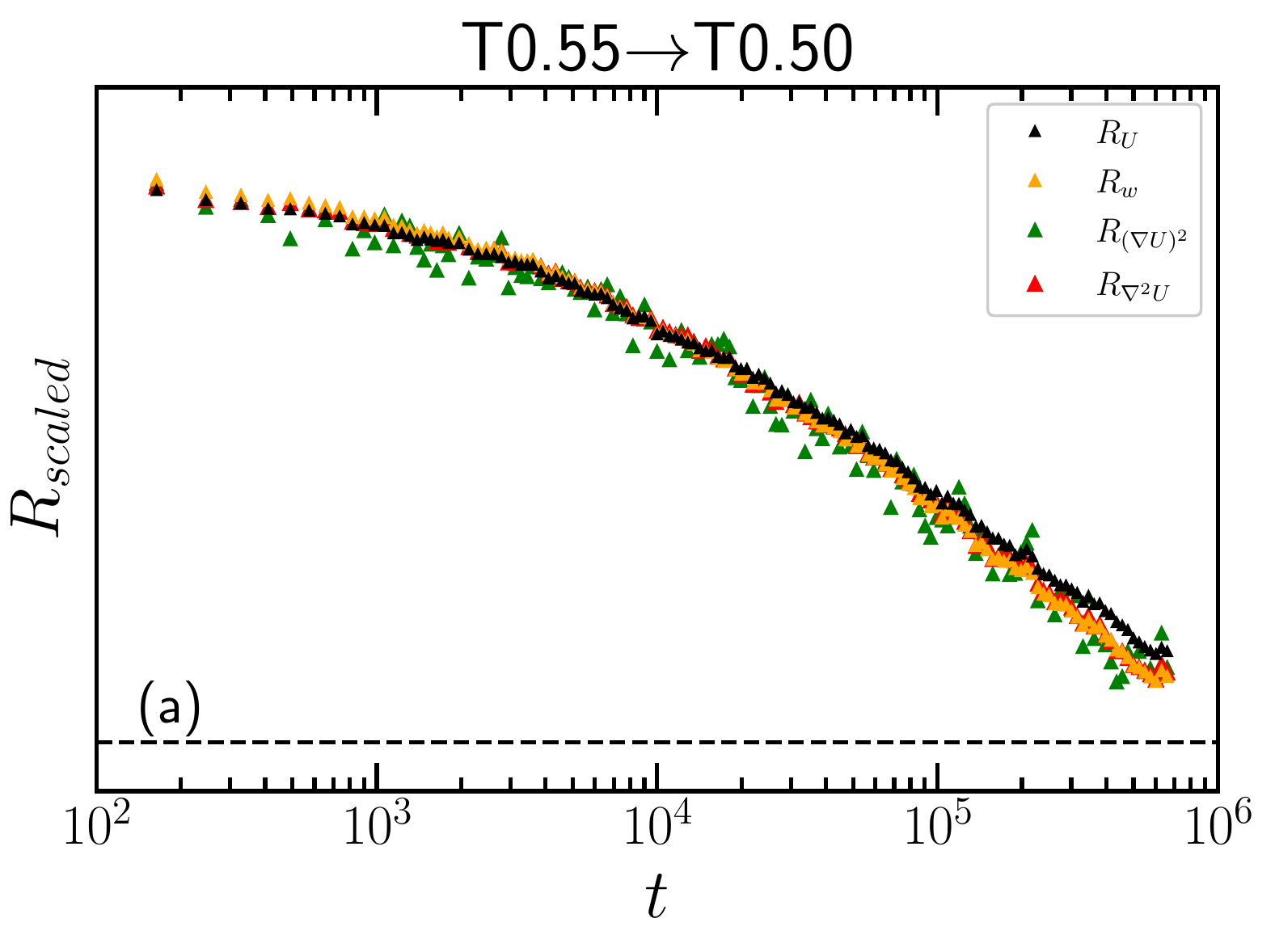}
	\includegraphics[width=7cm]{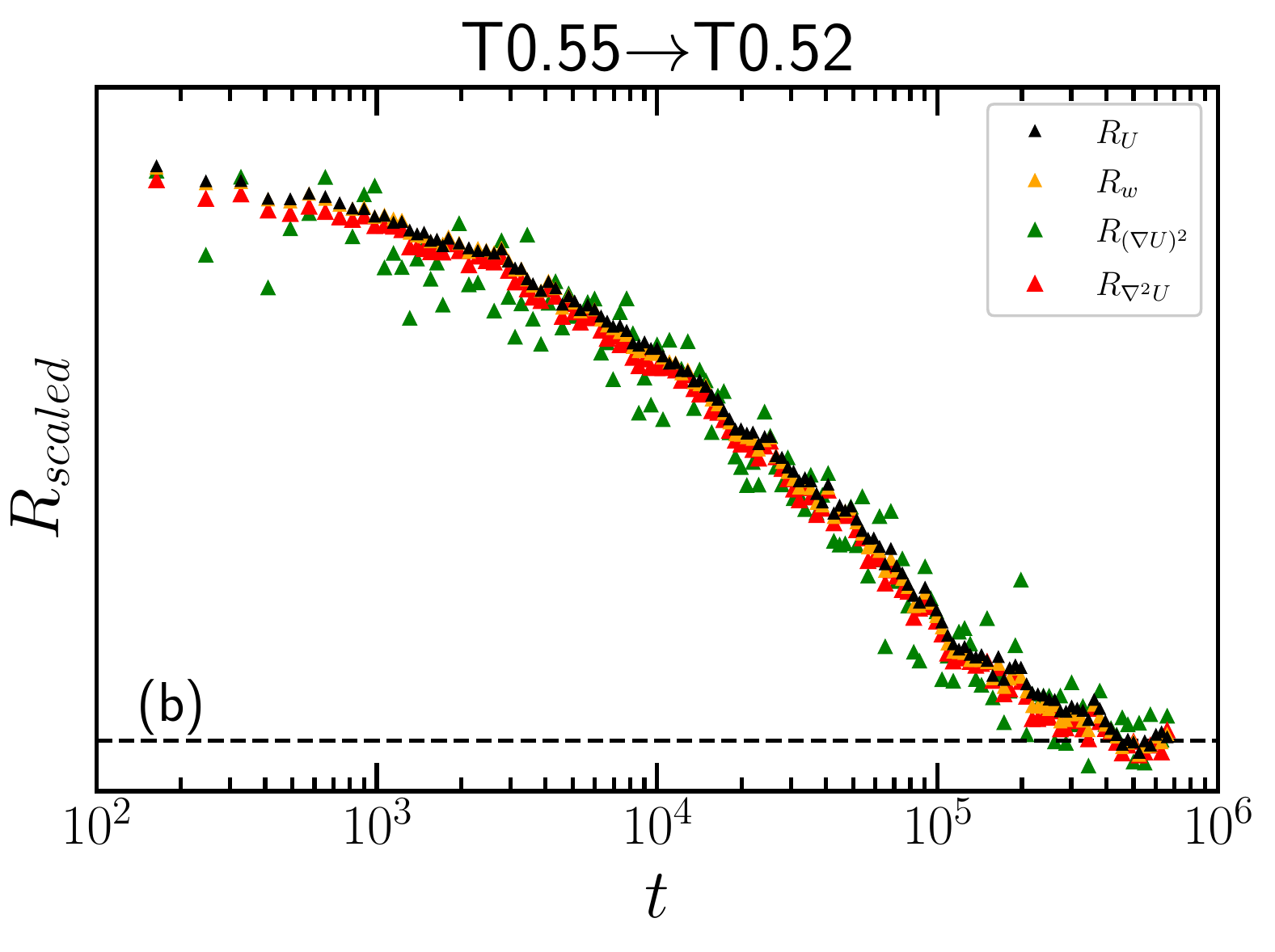}
	\includegraphics[width=7cm]{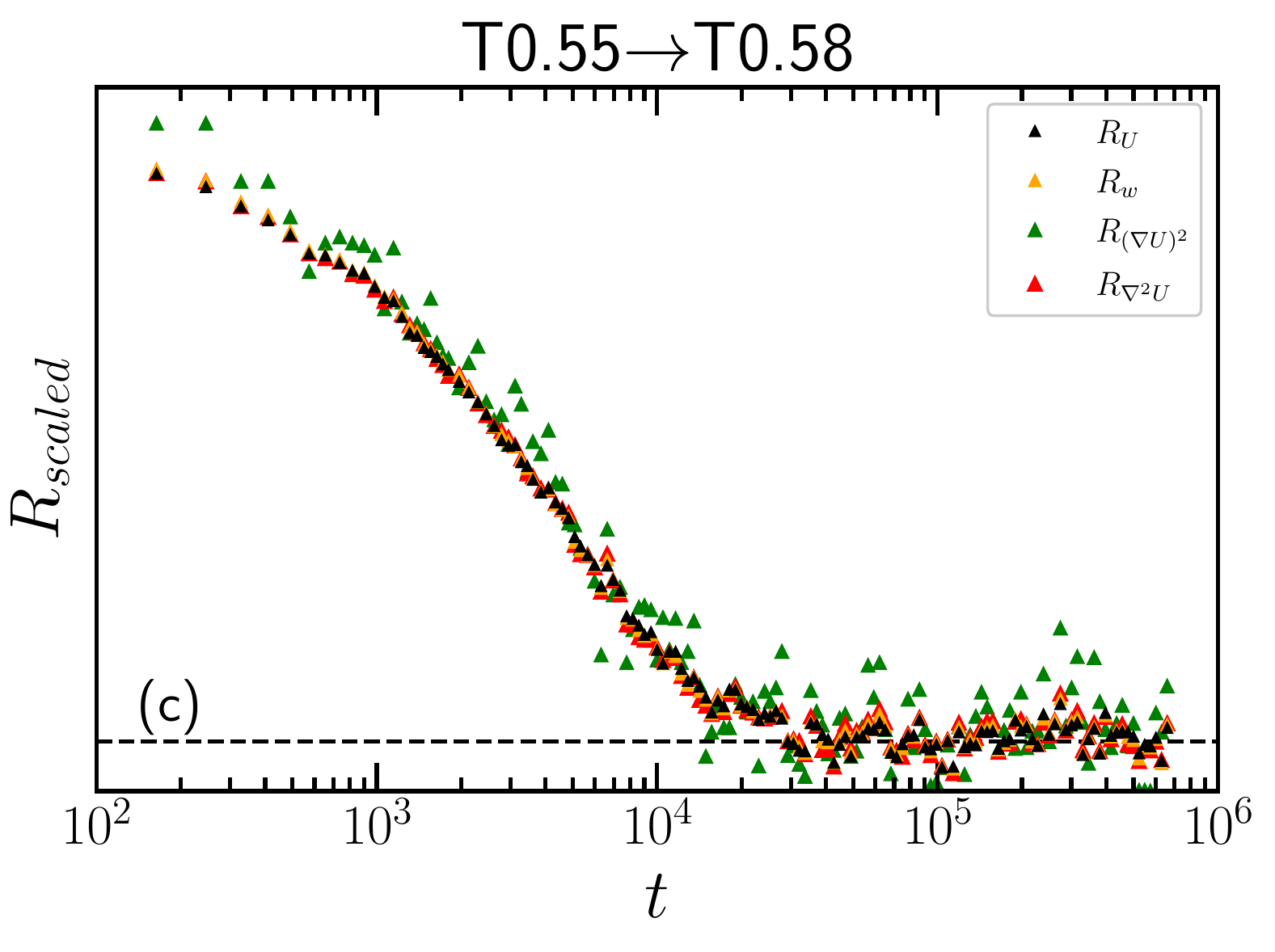}
	\includegraphics[width=7cm]{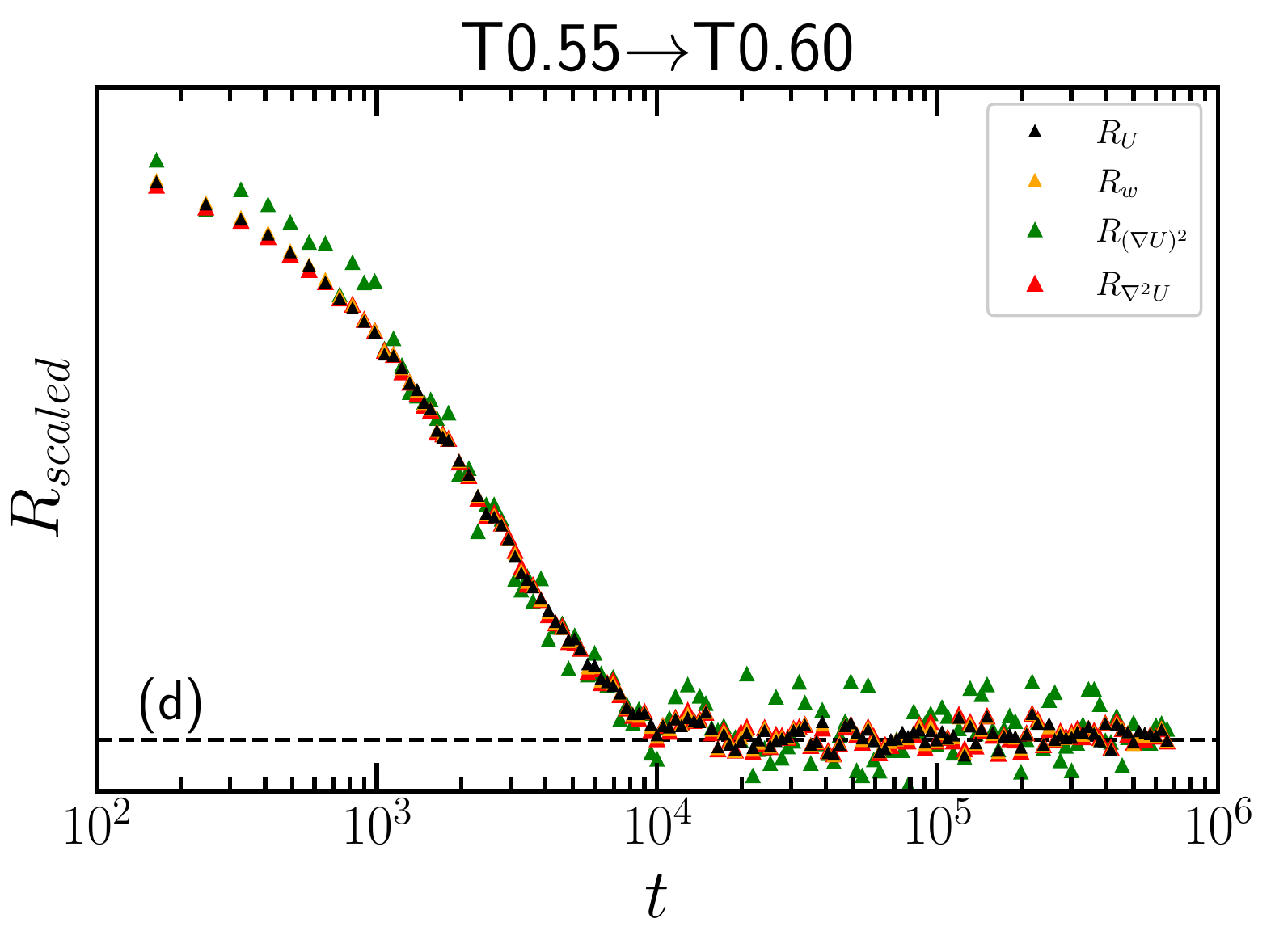}
	\caption{\label{relaxations_downjumps} Empirically scaled relaxation curves of the four quantities, plotted for each of the four jumps away from temperature $0.55$. The same colors codes are used here as in Fig. \ref{relaxations_upjumps}.}
\end{figure}

\section{Temperature-jump results}

Figure \ref{prediction1} investigates SPA for jumps to the same target temperature ($T_0=0.55$). In the upper panels, down jumps (green) were used to predict up jumps (blue), in the lower panels up jumps were used to predict down jumps. The relaxation curves do not start at unity because of the already mentioned ``instantaneous'' jump that occurs within the first few time steps of an aging simulation. The relative magnitude of this jump depends on the quantity in question. Jumps to different target temperatures were also investigated (Fig. \ref{prediction2}). The predictions in the upper panels are based on up jumps, while the predictions in the lower panels are based on down jumps. There are small deviations at the beginning and the predictions do not fit data as well for larger jumps as for smaller ones. Despite these minor deviations, we conclude that overall the results validate SPA for computer simulations. 

Figures \ref{relaxations_upjumps} and \ref{relaxations_downjumps} plot for each temperature jump all four relaxation curves. The curves have been scaled empirically by multiplying $R(t)$ by a constant in order to be able to compare the relaxing parts of the aging signals. We see that the four quantities relax almost identically. This demonstrates a physically appealing version of SPA according to which all four quantities age controlled by the same material time. After all, if the material time is to be thought of as reflecting the time on an ``internal clock'' of the aging system, the material time cannot depend on the quantity monitored. In the present context we note, however, that part of the scaled relaxing curves being virtually identical is not surprising. Thus it is known that binary Lennard-Jones systems have strong virial potential-energy correlations, implying that in equilibrium the virial is a linear function of the potential energy \cite{IV,boh12,ing15}. This extends to out-of-equilibrium situations \cite{III}. Thus one expects the virial and the potential energy to have the same relaxation functions, except for scaling constants. Likewise, the above observation that the configurational temperature equilibrates almost instantaneously implies that its numerator (the average squared force) and its denominator (the Laplacian of the potential energy) must be have the same relaxation functions.

\section{Discussion}

While physical aging is usually studied experimentally, computer simulations provide an alternative means for systematically investigating aging. For instance, it is much easier to control and rapidly change temperature in a computer simulation. It should be noted, though, that it is only with the presently available strong computing powers that it is possible to obtain simulation results of a quality that is comparable to that of aging experiments.

We find that SPA works well in computer simulations, albeit with an accuracy that decreases somewhat as the jump size increases. This is not surprising since a first-order Taylor expansion was used to derive SPA. It is important to note, however, that the largest temperature jumps considered here are, relatively, almost ten times larger than those of the experimental validations of the SPA formalism \cite{Hecksher2015,roed2019generalized} (10\% versus 1\%). Not surprisingly, the small deviations observed in some of the predictions at the beginning are larger, the larger the jumps are. Confirming previous findings by Powles and co-workers \cite{pow05}, we find that the configurational temperature, $k_BT_{\rm conf}$ does not age; on the other hand, its numerator and denominator both age following SPA. -- We recommend using data from up and down jumps with the same magnitude to the same target temperature when identifying $\chi_{\rm const}$ by use of Eq (\ref{param}). This is because by considering jumps with same target temperature, one avoids the need to model the temperature dependence of $\gamma_{\rm eq}$.

The above computer simulations were performed at constant volume. We have not attempted to test SPA in constant-pressure simulations, but expect that SPA applies equally well here. Thus the experiments confirming SPA were all performed at ambient pressure \cite{Hecksher2015,roed2019generalized}; moreover, the SPA derivation does not assume constant-volume conditions.

SPA is the simplest aging scenario consistent with the TN concept of a material time. This is because SPA is derived by assuming just a single relevant parameter and because first-order Taylor expansions are used \cite{Hecksher2015}. Our finding that all four quantities conform to SPA and relax in the same way shows that they are controlled by the same material time. This is not trivial. Whether all quantities of the binary LJ system age are controlled by this clock is an interesting question for future work. Also, it would be interesting to investigate whether the agreement with simulations may be improved by Taylor expanding to higher order, without making the SPA formalism highly involved or introducing a wealth of adjustable parameters.

\section*{Data availability}

The data of this study are available from the corresponding authors upon request.

\begin{acknowledgments}
This work was supported by the VILLUM Foundation's \textit{Matter} grant (16515).
\end{acknowledgments}


\begin{thebibliography}{58}%
	\makeatletter
	\providecommand \@ifxundefined [1]{%
		\@ifx{#1\undefined}
	}%
	\providecommand \@ifnum [1]{%
		\ifnum #1\expandafter \@firstoftwo
		\else \expandafter \@secondoftwo
		\fi
	}%
	\providecommand \@ifx [1]{%
		\ifx #1\expandafter \@firstoftwo
		\else \expandafter \@secondoftwo
		\fi
	}%
	\providecommand \natexlab [1]{#1}%
	\providecommand \enquote  [1]{``#1''}%
	\providecommand \bibnamefont  [1]{#1}%
	\providecommand \bibfnamefont [1]{#1}%
	\providecommand \citenamefont [1]{#1}%
	\providecommand \href@noop [0]{\@secondoftwo}%
	\providecommand \href [0]{\begingroup \@sanitize@url \@href}%
	\providecommand \@href[1]{\@@startlink{#1}\@@href}%
	\providecommand \@@href[1]{\endgroup#1\@@endlink}%
	\providecommand \@sanitize@url [0]{\catcode `\\12\catcode `\$12\catcode
		`\&12\catcode `\#12\catcode `\^12\catcode `\_12\catcode `\%12\relax}%
	\providecommand \@@startlink[1]{}%
	\providecommand \@@endlink[0]{}%
	\providecommand \url  [0]{\begingroup\@sanitize@url \@url }%
	\providecommand \@url [1]{\endgroup\@href {#1}{\urlprefix }}%
	\providecommand \urlprefix  [0]{URL }%
	\providecommand \Eprint [0]{\href }%
	\providecommand \doibase [0]{http://dx.doi.org/}%
	\providecommand \selectlanguage [0]{\@gobble}%
	\providecommand \bibinfo  [0]{\@secondoftwo}%
	\providecommand \bibfield  [0]{\@secondoftwo}%
	\providecommand \translation [1]{[#1]}%
	\providecommand \BibitemOpen [0]{}%
	\providecommand \bibitemStop [0]{}%
	\providecommand \bibitemNoStop [0]{.\EOS\space}%
	\providecommand \EOS [0]{\spacefactor3000\relax}%
	\providecommand \BibitemShut  [1]{\csname bibitem#1\endcsname}%
	\let\auto@bib@innerbib\@empty
	\bibitem [{\citenamefont {Cangialosi}(2014)}]{cangialosi2014dynamics}%
	\BibitemOpen
	\bibfield  {author} {\bibinfo {author} {\bibfnamefont {Daniele}\ \bibnamefont
			{Cangialosi}},\ }\bibfield  {title} {\enquote {\bibinfo {title} {Dynamics and
				thermodynamics of polymer glasses},}\ }\href@noop {} {\bibfield  {journal}
		{\bibinfo  {journal} {Journal of Physics: Condensed Matter}\ }\textbf
		{\bibinfo {volume} {26}},\ \bibinfo {pages} {153101} (\bibinfo {year}
		{2014})}\BibitemShut {NoStop}%
	\bibitem [{\citenamefont {Tool}(1946)}]{Tool1931}%
	\BibitemOpen
	\bibfield  {author} {\bibinfo {author} {\bibfnamefont {Arthur~Q.}\
			\bibnamefont {Tool}},\ }\bibfield  {title} {\enquote {\bibinfo {title}
			{Relation between inelastic deformability and thermal expansion of glass in
				its annealing range},}\ }\href {\doibase 10.1111/j.1151-2916.1946.tb11592.x}
	{\bibfield  {journal} {\bibinfo  {journal} {Journal of the American Ceramic
				Society}\ }\textbf {\bibinfo {volume} {29}},\ \bibinfo {pages} {240--253}
		(\bibinfo {year} {1946})}\BibitemShut {NoStop}%
	\bibitem [{\citenamefont {Narayanaswamy}(1971)}]{Narayana1971}%
	\BibitemOpen
	\bibfield  {author} {\bibinfo {author} {\bibfnamefont {O.~S.}\ \bibnamefont
			{Narayanaswamy}},\ }\bibfield  {title} {\enquote {\bibinfo {title} {A model
				of structural relaxation in glass},}\ }\href {\doibase
		10.1111/j.1151-2916.1971.tb12186.x} {\bibfield  {journal} {\bibinfo
			{journal} {Journal of the American Ceramic Society}\ }\textbf {\bibinfo
			{volume} {54}},\ \bibinfo {pages} {491--498} (\bibinfo {year}
		{1971})}\BibitemShut {NoStop}%
	\bibitem [{\citenamefont {Scherer}(1986)}]{scherer1986relaxation}%
	\BibitemOpen
	\bibfield  {author} {\bibinfo {author} {\bibfnamefont {G.~W.}\ \bibnamefont
			{Scherer}},\ }\href@noop {} {\emph {\bibinfo {title} {{Relaxation in Glass
					and Composites}}}}\ (\bibinfo  {publisher} {Wiley, New York},\ \bibinfo
	{year} {1986})\BibitemShut {NoStop}%
	\bibitem [{\citenamefont {Hodge}(1995)}]{Hodge1945}%
	\BibitemOpen
	\bibfield  {author} {\bibinfo {author} {\bibfnamefont {Ian~M.}\ \bibnamefont
			{Hodge}},\ }\bibfield  {title} {\enquote {\bibinfo {title} {Physical aging in
				polymer glasses},}\ }\href {\doibase 10.1126/science.267.5206.1945}
	{\bibfield  {journal} {\bibinfo  {journal} {Science}\ }\textbf {\bibinfo
			{volume} {267}},\ \bibinfo {pages} {1945--1947} (\bibinfo {year}
		{1995})}\BibitemShut {NoStop}%
	\bibitem [{\citenamefont {Moynihan}\ \emph {et~al.}(1976)\citenamefont
		{Moynihan}, \citenamefont {Easteal}, \citenamefont {De~Bolt},\ and\
		\citenamefont {Tucker}}]{moynihan1976dependence}%
	\BibitemOpen
	\bibfield  {author} {\bibinfo {author} {\bibfnamefont {Cornelius~T.}\
			\bibnamefont {Moynihan}}, \bibinfo {author} {\bibfnamefont {Allan~J.}\
			\bibnamefont {Easteal}}, \bibinfo {author} {\bibfnamefont {Mary~Ann}\
			\bibnamefont {De~Bolt}}, \ and\ \bibinfo {author} {\bibfnamefont {Joseph}\
			\bibnamefont {Tucker}},\ }\bibfield  {title} {\enquote {\bibinfo {title}
			{Dependence of the fictive temperature of glass on cooling rate},}\
	}\href@noop {} {\bibfield  {journal} {\bibinfo  {journal} {Journal of the
				American Ceramic Society}\ }\textbf {\bibinfo {volume} {59}},\ \bibinfo
		{pages} {12--16} (\bibinfo {year} {1976})}\BibitemShut {NoStop}%
	\bibitem [{\citenamefont {Olsen}\ \emph {et~al.}(1998)\citenamefont {Olsen},
		\citenamefont {Dyre},\ and\ \citenamefont
		{Christensen}}]{olsen1998structural}%
	\BibitemOpen
	\bibfield  {author} {\bibinfo {author} {\bibfnamefont {Niels~Boye}\
			\bibnamefont {Olsen}}, \bibinfo {author} {\bibfnamefont {Jeppe~C.}\
			\bibnamefont {Dyre}}, \ and\ \bibinfo {author} {\bibfnamefont {Tage}\
			\bibnamefont {Christensen}},\ }\bibfield  {title} {\enquote {\bibinfo {title}
			{Structural relaxation monitored by instantaneous shear modulus},}\
	}\href@noop {} {\bibfield  {journal} {\bibinfo  {journal} {Physical Review
				Letters}\ }\textbf {\bibinfo {volume} {81}},\ \bibinfo {pages} {1031}
		(\bibinfo {year} {1998})}\BibitemShut {NoStop}%
	\bibitem [{\citenamefont {Hecksher}\ \emph {et~al.}(2015)\citenamefont
		{Hecksher}, \citenamefont {Olsen},\ and\ \citenamefont
		{Dyre}}]{Hecksher2015}%
	\BibitemOpen
	\bibfield  {author} {\bibinfo {author} {\bibfnamefont {Tina}\ \bibnamefont
			{Hecksher}}, \bibinfo {author} {\bibfnamefont {Niels~Boye}\ \bibnamefont
			{Olsen}}, \ and\ \bibinfo {author} {\bibfnamefont {Jeppe~C.}\ \bibnamefont
			{Dyre}},\ }\bibfield  {title} {\enquote {\bibinfo {title} {Communication:
				Direct tests of single-parameter aging},}\ }\href {\doibase
		10.1063/1.4923000} {\bibfield  {journal} {\bibinfo  {journal} {The Journal of
				Chemical Physics}\ }\textbf {\bibinfo {volume} {142}},\ \bibinfo {pages}
		{241103} (\bibinfo {year} {2015})}\BibitemShut {NoStop}%
	\bibitem [{\citenamefont {Mauro}\ \emph {et~al.}(2009)\citenamefont {Mauro},
		\citenamefont {Loucks},\ and\ \citenamefont {Gupta}}]{mauro2009fictive}%
	\BibitemOpen
	\bibfield  {author} {\bibinfo {author} {\bibfnamefont {John~C.}\ \bibnamefont
			{Mauro}}, \bibinfo {author} {\bibfnamefont {Roger~J.}\ \bibnamefont
			{Loucks}}, \ and\ \bibinfo {author} {\bibfnamefont {Prabhat~K.}\ \bibnamefont
			{Gupta}},\ }\bibfield  {title} {\enquote {\bibinfo {title} {Fictive
				temperature and the glassy state},}\ }\href@noop {} {\bibfield  {journal}
		{\bibinfo  {journal} {Journal of the American Ceramic Society}\ }\textbf
		{\bibinfo {volume} {92}},\ \bibinfo {pages} {75--86} (\bibinfo {year}
		{2009})}\BibitemShut {NoStop}%
	\bibitem [{\citenamefont {Cugliandolo}\ and\ \citenamefont
		{Kurchan}(1994)}]{cugliandolo1994out}%
	\BibitemOpen
	\bibfield  {author} {\bibinfo {author} {\bibfnamefont {Leticia~F.}\
			\bibnamefont {Cugliandolo}}\ and\ \bibinfo {author} {\bibfnamefont {Jorge}\
			\bibnamefont {Kurchan}},\ }\bibfield  {title} {\enquote {\bibinfo {title} {On
				the out-of-equilibrium relaxation of the sherrington-kirkpatrick model},}\
	}\href@noop {} {\bibfield  {journal} {\bibinfo  {journal} {Journal of Physics
				A: Mathematical and General}\ }\textbf {\bibinfo {volume} {27}},\ \bibinfo
		{pages} {5749} (\bibinfo {year} {1994})}\BibitemShut {NoStop}%
	\bibitem [{\citenamefont {Kob}\ and\ \citenamefont
		{Barrat}(2000)}]{kob2000fluctuations}%
	\BibitemOpen
	\bibfield  {author} {\bibinfo {author} {\bibfnamefont {Walter}\ \bibnamefont
			{Kob}}\ and\ \bibinfo {author} {\bibfnamefont {J.-L.}\ \bibnamefont
			{Barrat}},\ }\bibfield  {title} {\enquote {\bibinfo {title} {Fluctuations,
				response and aging dynamics in a simple glass-forming liquid out of
				equilibrium},}\ }\href@noop {} {\bibfield  {journal} {\bibinfo  {journal}
			{The European Physical Journal B-Condensed Matter and Complex Systems}\
		}\textbf {\bibinfo {volume} {13}},\ \bibinfo {pages} {319--333} (\bibinfo
		{year} {2000})}\BibitemShut {NoStop}%
	\bibitem [{\citenamefont {Adolf}\ \emph {et~al.}(2007)\citenamefont {Adolf},
		\citenamefont {Chambers}, \citenamefont {Flemming}, \citenamefont {Budzien},\
		and\ \citenamefont {McCoy}}]{adolf2007potential}%
	\BibitemOpen
	\bibfield  {author} {\bibinfo {author} {\bibfnamefont {Douglas~B.}\
			\bibnamefont {Adolf}}, \bibinfo {author} {\bibfnamefont {Robert~S.}\
			\bibnamefont {Chambers}}, \bibinfo {author} {\bibfnamefont {Jesse}\
			\bibnamefont {Flemming}}, \bibinfo {author} {\bibfnamefont {Joanne}\
			\bibnamefont {Budzien}}, \ and\ \bibinfo {author} {\bibfnamefont {John}\
			\bibnamefont {McCoy}},\ }\bibfield  {title} {\enquote {\bibinfo {title}
			{Potential energy clock model: Justification and challenging predictions},}\
	}\href@noop {} {\bibfield  {journal} {\bibinfo  {journal} {Journal of
				Rheology}\ }\textbf {\bibinfo {volume} {51}},\ \bibinfo {pages} {517--540}
		(\bibinfo {year} {2007})}\BibitemShut {NoStop}%
	\bibitem [{\citenamefont {Castillo}\ and\ \citenamefont
		{Parsaeian}(2007)}]{castillo2007local}%
	\BibitemOpen
	\bibfield  {author} {\bibinfo {author} {\bibfnamefont {Horacio~E.}\
			\bibnamefont {Castillo}}\ and\ \bibinfo {author} {\bibfnamefont {Azita}\
			\bibnamefont {Parsaeian}},\ }\bibfield  {title} {\enquote {\bibinfo {title}
			{Local fluctuations in the ageing of a simple structural glass},}\
	}\href@noop {} {\bibfield  {journal} {\bibinfo  {journal} {Nature Physics}\
		}\textbf {\bibinfo {volume} {3}},\ \bibinfo {pages} {26--28} (\bibinfo {year}
		{2007})}\BibitemShut {NoStop}%
	\bibitem [{\citenamefont {Parsaeian}\ and\ \citenamefont
		{Castillo}(2009)}]{parsaeian2009equilibrium}%
	\BibitemOpen
	\bibfield  {author} {\bibinfo {author} {\bibfnamefont {Azita}\ \bibnamefont
			{Parsaeian}}\ and\ \bibinfo {author} {\bibfnamefont {Horacio~E.}\
			\bibnamefont {Castillo}},\ }\bibfield  {title} {\enquote {\bibinfo {title}
			{Equilibrium and nonequilibrium fluctuations in a glass-forming liquid},}\
	}\href@noop {} {\bibfield  {journal} {\bibinfo  {journal} {Physical Review
				Letters}\ }\textbf {\bibinfo {volume} {102}},\ \bibinfo {pages} {055704}
		(\bibinfo {year} {2009})}\BibitemShut {NoStop}%
	\bibitem [{\citenamefont {Kolvin}\ and\ \citenamefont
		{Bouchbinder}(2012)}]{kolvin2012simple}%
	\BibitemOpen
	\bibfield  {author} {\bibinfo {author} {\bibfnamefont {Itamar}\ \bibnamefont
			{Kolvin}}\ and\ \bibinfo {author} {\bibfnamefont {Eran}\ \bibnamefont
			{Bouchbinder}},\ }\bibfield  {title} {\enquote {\bibinfo {title} {Simple
				nonlinear equation for structural relaxation in glasses},}\ }\href@noop {}
	{\bibfield  {journal} {\bibinfo  {journal} {Physical Review E}\ }\textbf
		{\bibinfo {volume} {86}},\ \bibinfo {pages} {010501} (\bibinfo {year}
		{2012})}\BibitemShut {NoStop}%
	\bibitem [{\citenamefont {Struik}(1978)}]{Struik1978PhysicalAI}%
	\BibitemOpen
	\bibfield  {author} {\bibinfo {author} {\bibfnamefont {L.~C.~E.}\
			\bibnamefont {Struik}},\ }\bibfield  {title} {\enquote {\bibinfo {title}
			{Physical aging in amorphous polymers and other materials},}\ \ }(\bibinfo
	{publisher} {Elsevier, Amsterdam},\ \bibinfo {year} {1978})\BibitemShut
	{NoStop}%
	\bibitem [{\citenamefont {Hutchinson}(1995)}]{hutchinson1995prog}%
	\BibitemOpen
	\bibfield  {author} {\bibinfo {author} {\bibfnamefont {J.~M.}\ \bibnamefont
			{Hutchinson}},\ }\bibfield  {title} {\enquote {\bibinfo {title} {Physical
				aging of polymers},}\ }\href@noop {} {\bibfield  {journal} {\bibinfo
			{journal} {Prog. Polym. Sci.}\ }\textbf {\bibinfo {volume} {20}},\ \bibinfo
		{pages} {703--760} (\bibinfo {year} {1995})}\BibitemShut {NoStop}%
	\bibitem [{\citenamefont {Odegard}\ and\ \citenamefont
		{Bandyopadhyay}(2011)}]{odegard2011physical}%
	\BibitemOpen
	\bibfield  {author} {\bibinfo {author} {\bibfnamefont {G.~M.}\ \bibnamefont
			{Odegard}}\ and\ \bibinfo {author} {\bibfnamefont {A.}~\bibnamefont
			{Bandyopadhyay}},\ }\bibfield  {title} {\enquote {\bibinfo {title} {Physical
				aging of epoxy polymers and their composites},}\ }\href@noop {} {\bibfield
		{journal} {\bibinfo  {journal} {Journal of polymer science Part B: Polymer
				physics}\ }\textbf {\bibinfo {volume} {49}},\ \bibinfo {pages} {1695--1716}
		(\bibinfo {year} {2011})}\BibitemShut {NoStop}%
	\bibitem [{\citenamefont {Cangialosi}\ \emph {et~al.}(2013)\citenamefont
		{Cangialosi}, \citenamefont {Boucher}, \citenamefont {Alegr{\'\i}a},\ and\
		\citenamefont {Colmenero}}]{cangialosi2013physical}%
	\BibitemOpen
	\bibfield  {author} {\bibinfo {author} {\bibfnamefont {Daniele}\ \bibnamefont
			{Cangialosi}}, \bibinfo {author} {\bibfnamefont {Virginie~M.}\ \bibnamefont
			{Boucher}}, \bibinfo {author} {\bibfnamefont {Angel}\ \bibnamefont
			{Alegr{\'\i}a}}, \ and\ \bibinfo {author} {\bibfnamefont {Juan}\ \bibnamefont
			{Colmenero}},\ }\bibfield  {title} {\enquote {\bibinfo {title} {Physical
				aging in polymers and polymer nanocomposites: {R}ecent results and open
				questions},}\ }\href@noop {} {\bibfield  {journal} {\bibinfo  {journal} {Soft
				Matter}\ }\textbf {\bibinfo {volume} {9}},\ \bibinfo {pages} {8619--8630}
		(\bibinfo {year} {2013})}\BibitemShut {NoStop}%
	\bibitem [{\citenamefont {Grassia}\ and\ \citenamefont
		{Simon}(2012)}]{grassia2012modeling}%
	\BibitemOpen
	\bibfield  {author} {\bibinfo {author} {\bibfnamefont {Luigi}\ \bibnamefont
			{Grassia}}\ and\ \bibinfo {author} {\bibfnamefont {Sindee~L.}\ \bibnamefont
			{Simon}},\ }\bibfield  {title} {\enquote {\bibinfo {title} {Modeling volume
				relaxation of amorphous polymers: modification of the equation for the
				relaxation time in the kahr model},}\ }\href@noop {} {\bibfield  {journal}
		{\bibinfo  {journal} {Polymer}\ }\textbf {\bibinfo {volume} {53}},\ \bibinfo
		{pages} {3613--3620} (\bibinfo {year} {2012})}\BibitemShut {NoStop}%
	\bibitem [{\citenamefont {Qiao}\ and\ \citenamefont
		{Pelletier}(2014)}]{qiao2014dynamic}%
	\BibitemOpen
	\bibfield  {author} {\bibinfo {author} {\bibfnamefont {J.~C.}\ \bibnamefont
			{Qiao}}\ and\ \bibinfo {author} {\bibfnamefont {Jean-Marc}\ \bibnamefont
			{Pelletier}},\ }\bibfield  {title} {\enquote {\bibinfo {title} {Dynamic
				mechanical relaxation in bulk metallic glasses: a review},}\ }\href@noop {}
	{\bibfield  {journal} {\bibinfo  {journal} {Journal of Materials Science \&
				Technology}\ }\textbf {\bibinfo {volume} {30}},\ \bibinfo {pages} {523--545}
		(\bibinfo {year} {2014})}\BibitemShut {NoStop}%
	\bibitem [{\citenamefont {Song}\ \emph {et~al.}(2020)\citenamefont {Song},
		\citenamefont {Xu}, \citenamefont {Huo}, \citenamefont {Li}, \citenamefont
		{Wang}, \citenamefont {Ediger},\ and\ \citenamefont {Wang}}]{son20}%
	\BibitemOpen
	\bibfield  {author} {\bibinfo {author} {\bibfnamefont {L.}~\bibnamefont
			{Song}}, \bibinfo {author} {\bibfnamefont {W.}~\bibnamefont {Xu}}, \bibinfo
		{author} {\bibfnamefont {J.}~\bibnamefont {Huo}}, \bibinfo {author}
		{\bibfnamefont {F.}~\bibnamefont {Li}}, \bibinfo {author} {\bibfnamefont
			{L.-M.}\ \bibnamefont {Wang}}, \bibinfo {author} {\bibfnamefont {M.~D.}\
			\bibnamefont {Ediger}}, \ and\ \bibinfo {author} {\bibfnamefont {J.-Q.}\
			\bibnamefont {Wang}},\ }\bibfield  {title} {\enquote {\bibinfo {title}
			{Activation entropy as a key factor controlling the memory effect in
				glasses},}\ }\href {\doibase 10.1103/PhysRevLett.125.135501} {\bibfield
		{journal} {\bibinfo  {journal} {Phys. Rev. Lett.}\ }\textbf {\bibinfo
			{volume} {125}},\ \bibinfo {pages} {135501} (\bibinfo {year}
		{2020})}\BibitemShut {NoStop}%
	\bibitem [{\citenamefont {Lundgren}\ \emph {et~al.}(1983)\citenamefont
		{Lundgren}, \citenamefont {Svedlindh}, \citenamefont {Nordblad},\ and\
		\citenamefont {Beckman}}]{lundgren1983dynamics}%
	\BibitemOpen
	\bibfield  {author} {\bibinfo {author} {\bibfnamefont {L.}~\bibnamefont
			{Lundgren}}, \bibinfo {author} {\bibfnamefont {P.}~\bibnamefont {Svedlindh}},
		\bibinfo {author} {\bibfnamefont {P.}~\bibnamefont {Nordblad}}, \ and\
		\bibinfo {author} {\bibfnamefont {O.}~\bibnamefont {Beckman}},\ }\bibfield
	{title} {\enquote {\bibinfo {title} {Dynamics of the relaxation-time spectrum
				in a cumn spin-glass},}\ }\href@noop {} {\bibfield  {journal} {\bibinfo
			{journal} {Physical Review Letters}\ }\textbf {\bibinfo {volume} {51}},\
		\bibinfo {pages} {911} (\bibinfo {year} {1983})}\BibitemShut {NoStop}%
	\bibitem [{\citenamefont {Berthier}\ and\ \citenamefont
		{Bouchaud}(2002)}]{berthier2002geometrical}%
	\BibitemOpen
	\bibfield  {author} {\bibinfo {author} {\bibfnamefont {Ludovic}\ \bibnamefont
			{Berthier}}\ and\ \bibinfo {author} {\bibfnamefont {Jean-Philippe}\
			\bibnamefont {Bouchaud}},\ }\bibfield  {title} {\enquote {\bibinfo {title}
			{Geometrical aspects of aging and rejuvenation in the ising spin glass: A
				numerical study},}\ }\href@noop {} {\bibfield  {journal} {\bibinfo  {journal}
			{Physical Review B}\ }\textbf {\bibinfo {volume} {66}},\ \bibinfo {pages}
		{054404} (\bibinfo {year} {2002})}\BibitemShut {NoStop}%
	\bibitem [{\citenamefont {Spinner}\ and\ \citenamefont
		{Napolitano}(1966)}]{spinner1966further}%
	\BibitemOpen
	\bibfield  {author} {\bibinfo {author} {\bibfnamefont {Sam}\ \bibnamefont
			{Spinner}}\ and\ \bibinfo {author} {\bibfnamefont {Albert}\ \bibnamefont
			{Napolitano}},\ }\bibfield  {title} {\enquote {\bibinfo {title} {Further
				studies in the annealing of a borosilicate glass},}\ }\href@noop {}
	{\bibfield  {journal} {\bibinfo  {journal} {Journal of Research of the
				National Bureau of Standards. Section A, Physics and Chemistry}\ }\textbf
		{\bibinfo {volume} {70}},\ \bibinfo {pages} {147} (\bibinfo {year}
		{1966})}\BibitemShut {NoStop}%
	\bibitem [{\citenamefont {Schlosser}\ and\ \citenamefont
		{Sch{\"o}nhals}(1991)}]{schlosser1991dielectric}%
	\BibitemOpen
	\bibfield  {author} {\bibinfo {author} {\bibfnamefont {E.}~\bibnamefont
			{Schlosser}}\ and\ \bibinfo {author} {\bibfnamefont {A.}~\bibnamefont
			{Sch{\"o}nhals}},\ }\bibfield  {title} {\enquote {\bibinfo {title}
			{Dielectric relaxation during physical ageing},}\ }\href@noop {} {\bibfield
		{journal} {\bibinfo  {journal} {Polymer}\ }\textbf {\bibinfo {volume} {32}},\
		\bibinfo {pages} {2135--2140} (\bibinfo {year} {1991})}\BibitemShut {NoStop}%
	\bibitem [{\citenamefont {Leheny}\ and\ \citenamefont
		{Nagel}(1998)}]{leheny1998frequency}%
	\BibitemOpen
	\bibfield  {author} {\bibinfo {author} {\bibfnamefont {Robert~L.}\
			\bibnamefont {Leheny}}\ and\ \bibinfo {author} {\bibfnamefont {Sidney~R.}\
			\bibnamefont {Nagel}},\ }\bibfield  {title} {\enquote {\bibinfo {title}
			{Frequency-domain study of physical aging in a simple liquid},}\ }\href@noop
	{} {\bibfield  {journal} {\bibinfo  {journal} {Physical Review B}\ }\textbf
		{\bibinfo {volume} {57}},\ \bibinfo {pages} {5154} (\bibinfo {year}
		{1998})}\BibitemShut {NoStop}%
	\bibitem [{\citenamefont {Lunkenheimer}\ \emph {et~al.}(2005)\citenamefont
		{Lunkenheimer}, \citenamefont {Wehn}, \citenamefont {Schneider},\ and\
		\citenamefont {Loidl}}]{lunkenheimer2005glassy}%
	\BibitemOpen
	\bibfield  {author} {\bibinfo {author} {\bibfnamefont {Peter}\ \bibnamefont
			{Lunkenheimer}}, \bibinfo {author} {\bibfnamefont {Robert}\ \bibnamefont
			{Wehn}}, \bibinfo {author} {\bibfnamefont {Ulrich}\ \bibnamefont
			{Schneider}}, \ and\ \bibinfo {author} {\bibfnamefont {Alois}\ \bibnamefont
			{Loidl}},\ }\bibfield  {title} {\enquote {\bibinfo {title} {Glassy aging
				dynamics},}\ }\href@noop {} {\bibfield  {journal} {\bibinfo  {journal}
			{Physical Review Letters}\ }\textbf {\bibinfo {volume} {95}},\ \bibinfo
		{pages} {055702} (\bibinfo {year} {2005})}\BibitemShut {NoStop}%
	\bibitem [{\citenamefont {Richert}(2015)}]{richert2015supercooled}%
	\BibitemOpen
	\bibfield  {author} {\bibinfo {author} {\bibfnamefont {Ranko}\ \bibnamefont
			{Richert}},\ }\bibfield  {title} {\enquote {\bibinfo {title} {Supercooled
				liquids and glasses by dielectric relaxation spectroscopy},}\ }\href@noop {}
	{\bibfield  {journal} {\bibinfo  {journal} {Adv. Chem. Phys.}\ }\textbf
		{\bibinfo {volume} {156}},\ \bibinfo {pages} {101--195} (\bibinfo {year}
		{2015})}\BibitemShut {NoStop}%
	\bibitem [{\citenamefont {Hecksher}\ \emph {et~al.}(2010)\citenamefont
		{Hecksher}, \citenamefont {Olsen}, \citenamefont {Niss},\ and\ \citenamefont
		{Dyre}}]{hecksher2010physical}%
	\BibitemOpen
	\bibfield  {author} {\bibinfo {author} {\bibfnamefont {Tina}\ \bibnamefont
			{Hecksher}}, \bibinfo {author} {\bibfnamefont {Niels~Boye}\ \bibnamefont
			{Olsen}}, \bibinfo {author} {\bibfnamefont {Kristine}\ \bibnamefont {Niss}},
		\ and\ \bibinfo {author} {\bibfnamefont {Jeppe~C.}\ \bibnamefont {Dyre}},\
	}\bibfield  {title} {\enquote {\bibinfo {title} {Physical aging of molecular
				glasses studied by a device allowing for rapid thermal equilibration},}\
	}\href@noop {} {\bibfield  {journal} {\bibinfo  {journal} {The Journal of
				Chemical Physics}\ }\textbf {\bibinfo {volume} {133}},\ \bibinfo {pages}
		{174514} (\bibinfo {year} {2010})}\BibitemShut {NoStop}%
	\bibitem [{\citenamefont {Wehn}\ \emph {et~al.}(2007)\citenamefont {Wehn},
		\citenamefont {Lunkenheimer},\ and\ \citenamefont
		{Loidl}}]{wehn2007broadband}%
	\BibitemOpen
	\bibfield  {author} {\bibinfo {author} {\bibfnamefont {R.}~\bibnamefont
			{Wehn}}, \bibinfo {author} {\bibfnamefont {Peter}\ \bibnamefont
			{Lunkenheimer}}, \ and\ \bibinfo {author} {\bibfnamefont {Alois}\
			\bibnamefont {Loidl}},\ }\bibfield  {title} {\enquote {\bibinfo {title}
			{Broadband dielectric spectroscopy and aging of glass formers},}\ }\href@noop
	{} {\bibfield  {journal} {\bibinfo  {journal} {Journal of Non-Crystalline
				Solids}\ }\textbf {\bibinfo {volume} {353}},\ \bibinfo {pages} {3862--3870}
		(\bibinfo {year} {2007})}\BibitemShut {NoStop}%
	\bibitem [{\citenamefont {Dyre}\ and\ \citenamefont
		{Olsen}(2003)}]{dyre2003minimal}%
	\BibitemOpen
	\bibfield  {author} {\bibinfo {author} {\bibfnamefont {Jeppe~C.}\
			\bibnamefont {Dyre}}\ and\ \bibinfo {author} {\bibfnamefont {Niels~Boye}\
			\bibnamefont {Olsen}},\ }\bibfield  {title} {\enquote {\bibinfo {title}
			{Minimal model for beta relaxation in viscous liquids},}\ }\href@noop {}
	{\bibfield  {journal} {\bibinfo  {journal} {Physical Review Letters}\
		}\textbf {\bibinfo {volume} {91}},\ \bibinfo {pages} {155703} (\bibinfo
		{year} {2003})}\BibitemShut {NoStop}%
	\bibitem [{\citenamefont {Kovacs}(1963)}]{kov63}%
	\BibitemOpen
	\bibfield  {author} {\bibinfo {author} {\bibfnamefont {A.~J.}\ \bibnamefont
			{Kovacs}},\ }\bibfield  {title} {\enquote {\bibinfo {title} {Transition
				vitreuse dans les polymeres amorphes. {Etude} phenomenologique},}\
	}\href@noop {} {\bibfield  {journal} {\bibinfo  {journal} {Fortschr.
				Hochpolym.-Forsch.}\ }\textbf {\bibinfo {volume} {3}},\ \bibinfo {pages}
		{394--507} (\bibinfo {year} {1963})}\BibitemShut {NoStop}%
	\bibitem [{\citenamefont {Di}\ \emph {et~al.}(2011)\citenamefont {Di},
		\citenamefont {Win}, \citenamefont {McKenna}, \citenamefont {Narita},
		\citenamefont {Lequeux}, \citenamefont {Pullela},\ and\ \citenamefont
		{Cheng}}]{di11}%
	\BibitemOpen
	\bibfield  {author} {\bibinfo {author} {\bibfnamefont {X.}~\bibnamefont
			{Di}}, \bibinfo {author} {\bibfnamefont {K.~Z.}\ \bibnamefont {Win}},
		\bibinfo {author} {\bibfnamefont {G.~B.}\ \bibnamefont {McKenna}}, \bibinfo
		{author} {\bibfnamefont {T.}~\bibnamefont {Narita}}, \bibinfo {author}
		{\bibfnamefont {F.}~\bibnamefont {Lequeux}}, \bibinfo {author} {\bibfnamefont
			{S.~R.}\ \bibnamefont {Pullela}}, \ and\ \bibinfo {author} {\bibfnamefont
			{Z.}~\bibnamefont {Cheng}},\ }\bibfield  {title} {\enquote {\bibinfo {title}
			{Signatures of structural recovery in colloidal glasses},}\ }\href {\doibase
		10.1103/PhysRevLett.106.095701} {\bibfield  {journal} {\bibinfo  {journal}
			{Phys. Rev. Lett.}\ }\textbf {\bibinfo {volume} {106}},\ \bibinfo {pages}
		{095701} (\bibinfo {year} {2011})}\BibitemShut {NoStop}%
	\bibitem [{\citenamefont {McKenna}\ and\ \citenamefont {Simon}(2017)}]{mck17}%
	\BibitemOpen
	\bibfield  {author} {\bibinfo {author} {\bibfnamefont {G.~B.}\ \bibnamefont
			{McKenna}}\ and\ \bibinfo {author} {\bibfnamefont {S.~L.}\ \bibnamefont
			{Simon}},\ }\bibfield  {title} {\enquote {\bibinfo {title} {50th anniversary
				perspective: {Challenges} in the dynamics and kinetics of glass-forming
				polymers},}\ }\href {\doibase 10.1021/acs.macromol.7b01014} {\bibfield
		{journal} {\bibinfo  {journal} {Macromolecules}\ }\textbf {\bibinfo {volume}
			{50}},\ \bibinfo {pages} {6333--6361} (\bibinfo {year} {2017})}\BibitemShut
	{NoStop}%
	\bibitem [{\citenamefont {Mazurin}(1977)}]{mazurin1977relaxation}%
	\BibitemOpen
	\bibfield  {author} {\bibinfo {author} {\bibfnamefont {O.~V.}\ \bibnamefont
			{Mazurin}},\ }\bibfield  {title} {\enquote {\bibinfo {title} {Relaxation
				phenomena in glass},}\ }\href@noop {} {\bibfield  {journal} {\bibinfo
			{journal} {Journal of Non-Crystalline Solids}\ }\textbf {\bibinfo {volume}
			{25}},\ \bibinfo {pages} {129--169} (\bibinfo {year} {1977})}\BibitemShut
	{NoStop}%
	\bibitem [{\citenamefont {McKenna}(1994)}]{mckenna1994physics}%
	\BibitemOpen
	\bibfield  {author} {\bibinfo {author} {\bibfnamefont {Gregory~B.}\
			\bibnamefont {McKenna}},\ }\bibfield  {title} {\enquote {\bibinfo {title} {On
				the physics required for prediction of long term performance of polymers and
				their composites},}\ }\href@noop {} {\bibfield  {journal} {\bibinfo
			{journal} {J. Res. Natl. Inst. Stand. Technol.}\ }\textbf {\bibinfo {volume}
			{99}},\ \bibinfo {pages} {169--169} (\bibinfo {year} {1994})}\BibitemShut
	{NoStop}%
	\bibitem [{\citenamefont {Hodge}(1994)}]{hodge1994enthalpy}%
	\BibitemOpen
	\bibfield  {author} {\bibinfo {author} {\bibfnamefont {Ian~M}\ \bibnamefont
			{Hodge}},\ }\bibfield  {title} {\enquote {\bibinfo {title} {Enthalpy
				relaxation and recovery in amorphous materials},}\ }\href@noop {} {\bibfield
		{journal} {\bibinfo  {journal} {Journal of Non-Crystalline Solids}\ }\textbf
		{\bibinfo {volume} {169}},\ \bibinfo {pages} {211--266} (\bibinfo {year}
		{1994})}\BibitemShut {NoStop}%
	\bibitem [{\citenamefont {Avramov}(1996)}]{avramov1996kinetics}%
	\BibitemOpen
	\bibfield  {author} {\bibinfo {author} {\bibfnamefont {I.}~\bibnamefont
			{Avramov}},\ }\bibfield  {title} {\enquote {\bibinfo {title} {Kinetics of
				structural relaxation of glass-forming melts},}\ }\href@noop {} {\bibfield
		{journal} {\bibinfo  {journal} {Thermochimica Acta}\ }\textbf {\bibinfo
			{volume} {280}},\ \bibinfo {pages} {363--382} (\bibinfo {year}
		{1996})}\BibitemShut {NoStop}%
	\bibitem [{\citenamefont {Roed}\ \emph {et~al.}(2019)\citenamefont {Roed},
		\citenamefont {Hecksher}, \citenamefont {Dyre},\ and\ \citenamefont
		{Niss}}]{roed2019generalized}%
	\BibitemOpen
	\bibfield  {author} {\bibinfo {author} {\bibfnamefont {Lisa~Anita}\
			\bibnamefont {Roed}}, \bibinfo {author} {\bibfnamefont {Tina}\ \bibnamefont
			{Hecksher}}, \bibinfo {author} {\bibfnamefont {Jeppe~C.}\ \bibnamefont
			{Dyre}}, \ and\ \bibinfo {author} {\bibfnamefont {Kristine}\ \bibnamefont
			{Niss}},\ }\bibfield  {title} {\enquote {\bibinfo {title} {Generalized
				single-parameter aging tests and their application to glycerol},}\
	}\href@noop {} {\bibfield  {journal} {\bibinfo  {journal} {The Journal of
				Chemical Physics}\ }\textbf {\bibinfo {volume} {150}},\ \bibinfo {pages}
		{044501} (\bibinfo {year} {2019})}\BibitemShut {NoStop}%
	\bibitem [{\citenamefont {Dyre}(2006)}]{dyr06}%
	\BibitemOpen
	\bibfield  {author} {\bibinfo {author} {\bibfnamefont {J.~C.}\ \bibnamefont
			{Dyre}},\ }\bibfield  {title} {\enquote {\bibinfo {title} {The glass
				transition and elastic models of glass-forming liquids},}\ }\href@noop {}
	{\bibfield  {journal} {\bibinfo  {journal} {Rev. Mod. Phys.}\ }\textbf
		{\bibinfo {volume} {78}},\ \bibinfo {pages} {953--972} (\bibinfo {year}
		{2006})}\BibitemShut {NoStop}%
	\bibitem [{\citenamefont {Ritland}(1956)}]{ritland1956limitations}%
	\BibitemOpen
	\bibfield  {author} {\bibinfo {author} {\bibfnamefont {H.~N.}\ \bibnamefont
			{Ritland}},\ }\bibfield  {title} {\enquote {\bibinfo {title} {Limitations of
				the fictive temperature concept},}\ }\href@noop {} {\bibfield  {journal}
		{\bibinfo  {journal} {Journal of the American Ceramic Society}\ }\textbf
		{\bibinfo {volume} {39}},\ \bibinfo {pages} {403--406} (\bibinfo {year}
		{1956})}\BibitemShut {NoStop}%
	\bibitem [{\citenamefont {Dyre}(2015)}]{dyre2015narayanaswamy}%
	\BibitemOpen
	\bibfield  {author} {\bibinfo {author} {\bibfnamefont {Jeppe~C.}\
			\bibnamefont {Dyre}},\ }\bibfield  {title} {\enquote {\bibinfo {title}
			{Narayanaswamy’s 1971 aging theory and material time},}\ }\href@noop {}
	{\bibfield  {journal} {\bibinfo  {journal} {The Journal of Chemical Physics}\
		}\textbf {\bibinfo {volume} {143}},\ \bibinfo {pages} {114507} (\bibinfo
		{year} {2015})}\BibitemShut {NoStop}%
	\bibitem [{\citenamefont {Kob}\ and\ \citenamefont {Andersen}(1995)}]{ka1}%
	\BibitemOpen
	\bibfield  {author} {\bibinfo {author} {\bibfnamefont {W.}~\bibnamefont
			{Kob}}\ and\ \bibinfo {author} {\bibfnamefont {H.~C.}\ \bibnamefont
			{Andersen}},\ }\bibfield  {title} {\enquote {\bibinfo {title} {{Testing
					Mode-Coupling Theory for a Supercooled Binary Lennard-Jones Mixture I: The
					van Hove Correlation Function}},}\ }\href@noop {} {\bibfield  {journal}
		{\bibinfo  {journal} {Phys. Rev. E}\ }\textbf {\bibinfo {volume} {51}},\
		\bibinfo {pages} {4626--4641} (\bibinfo {year} {1995})}\BibitemShut {NoStop}%
	\bibitem [{\citenamefont {Vollmayr}\ \emph {et~al.}(1996)\citenamefont
		{Vollmayr}, \citenamefont {Kob},\ and\ \citenamefont {Binder}}]{vol96}%
	\BibitemOpen
	\bibfield  {author} {\bibinfo {author} {\bibfnamefont {K.}~\bibnamefont
			{Vollmayr}}, \bibinfo {author} {\bibfnamefont {W.}~\bibnamefont {Kob}}, \
		and\ \bibinfo {author} {\bibfnamefont {K.}~\bibnamefont {Binder}},\
	}\bibfield  {title} {\enquote {\bibinfo {title} {How do the properties of a
				glass depend on the cooling rate? {A} computer simulation study of a
				{Lennard‐Jones} system},}\ }\href {\doibase 10.1063/1.472326} {\bibfield
		{journal} {\bibinfo  {journal} {J. Chem. Phys.}\ }\textbf {\bibinfo {volume}
			{105}},\ \bibinfo {pages} {4714--4728} (\bibinfo {year} {1996})}\BibitemShut
	{NoStop}%
	\bibitem [{\citenamefont {Parsaeian}\ and\ \citenamefont
		{Castillo}(2008)}]{par08a}%
	\BibitemOpen
	\bibfield  {author} {\bibinfo {author} {\bibfnamefont {A.}~\bibnamefont
			{Parsaeian}}\ and\ \bibinfo {author} {\bibfnamefont {H.~E.}\ \bibnamefont
			{Castillo}},\ }\bibfield  {title} {\enquote {\bibinfo {title} {Growth of
				spatial correlations in the aging of a simple structural glass},}\ }\href
	{\doibase 10.1103/PhysRevE.78.060105} {\bibfield  {journal} {\bibinfo
			{journal} {Phys. Rev. E}\ }\textbf {\bibinfo {volume} {78}},\ \bibinfo
		{pages} {060105} (\bibinfo {year} {2008})}\BibitemShut {NoStop}%
	\bibitem [{\citenamefont {Rehwald}\ \emph {et~al.}(2010)\citenamefont
		{Rehwald}, \citenamefont {Gnan}, \citenamefont {Heuer}, \citenamefont
		{Schr\o{}der}, \citenamefont {Dyre},\ and\ \citenamefont
		{Diezemann}}]{reh10}%
	\BibitemOpen
	\bibfield  {author} {\bibinfo {author} {\bibfnamefont {C.}~\bibnamefont
			{Rehwald}}, \bibinfo {author} {\bibfnamefont {N.}~\bibnamefont {Gnan}},
		\bibinfo {author} {\bibfnamefont {A.}~\bibnamefont {Heuer}}, \bibinfo
		{author} {\bibfnamefont {T.}~\bibnamefont {Schr\o{}der}}, \bibinfo {author}
		{\bibfnamefont {J.~C.}\ \bibnamefont {Dyre}}, \ and\ \bibinfo {author}
		{\bibfnamefont {G.}~\bibnamefont {Diezemann}},\ }\bibfield  {title} {\enquote
		{\bibinfo {title} {Aging effects manifested in the potential-energy landscape
				of a model glass former},}\ }\href {\doibase 10.1103/PhysRevE.82.021503}
	{\bibfield  {journal} {\bibinfo  {journal} {Phys. Rev. E}\ }\textbf {\bibinfo
			{volume} {82}},\ \bibinfo {pages} {021503} (\bibinfo {year}
		{2010})}\BibitemShut {NoStop}%
	\bibitem [{\citenamefont {Gnan}\ \emph {et~al.}(2013)\citenamefont {Gnan},
		\citenamefont {Maggi}, \citenamefont {Parisi},\ and\ \citenamefont
		{Sciortino}}]{gna13}%
	\BibitemOpen
	\bibfield  {author} {\bibinfo {author} {\bibfnamefont {N.}~\bibnamefont
			{Gnan}}, \bibinfo {author} {\bibfnamefont {C.}~\bibnamefont {Maggi}},
		\bibinfo {author} {\bibfnamefont {G.}~\bibnamefont {Parisi}}, \ and\ \bibinfo
		{author} {\bibfnamefont {F.}~\bibnamefont {Sciortino}},\ }\bibfield  {title}
	{\enquote {\bibinfo {title} {Generalized fluctuation-dissipation relation and
				effective temperature upon heating a deeply supercooled liquid},}\ }\href
	{\doibase 10.1103/PhysRevLett.110.035701} {\bibfield  {journal} {\bibinfo
			{journal} {Phys. Rev. Lett.}\ }\textbf {\bibinfo {volume} {110}},\ \bibinfo
		{pages} {035701} (\bibinfo {year} {2013})}\BibitemShut {NoStop}%
	\bibitem [{\citenamefont {Priezjev}(2018)}]{pri18}%
	\BibitemOpen
	\bibfield  {author} {\bibinfo {author} {\bibfnamefont {N.~V.}\ \bibnamefont
			{Priezjev}},\ }\bibfield  {title} {\enquote {\bibinfo {title} {Slow
				relaxation dynamics in binary glasses during stress-controlled,
				tension-compression cyclic loading},}\ }\href {\doibase
		10.1016/j.commatsci.2018.06.044} {\bibfield  {journal} {\bibinfo  {journal}
			{Computational Materials Science}\ }\textbf {\bibinfo {volume} {153}},\
		\bibinfo {pages} {235 -- 240} (\bibinfo {year} {2018})}\BibitemShut {NoStop}%
	\bibitem [{\citenamefont {Ingebrigtsen}\ \emph {et~al.}(2019)\citenamefont
		{Ingebrigtsen}, \citenamefont {Dyre}, \citenamefont {Schr\o{}der},\ and\
		\citenamefont {Royall}}]{PhysRevX.9.031016}%
	\BibitemOpen
	\bibfield  {author} {\bibinfo {author} {\bibfnamefont {Trond~S.}\
			\bibnamefont {Ingebrigtsen}}, \bibinfo {author} {\bibfnamefont {Jeppe~C.}\
			\bibnamefont {Dyre}}, \bibinfo {author} {\bibfnamefont {Thomas~B.}\
			\bibnamefont {Schr\o{}der}}, \ and\ \bibinfo {author} {\bibfnamefont
			{C.~Patrick}\ \bibnamefont {Royall}},\ }\bibfield  {title} {\enquote
		{\bibinfo {title} {Crystallization instability in glass-forming mixtures},}\
	}\href {\doibase 10.1103/PhysRevX.9.031016} {\bibfield  {journal} {\bibinfo
			{journal} {Phys. Rev. X}\ }\textbf {\bibinfo {volume} {9}},\ \bibinfo {pages}
		{031016} (\bibinfo {year} {2019})}\BibitemShut {NoStop}%
	\bibitem [{\citenamefont {Pedersen}\ \emph {et~al.}(2018)\citenamefont
		{Pedersen}, \citenamefont {Schr{\o}der},\ and\ \citenamefont {Dyre}}]{ped18}%
	\BibitemOpen
	\bibfield  {author} {\bibinfo {author} {\bibfnamefont {U.~R.}\ \bibnamefont
			{Pedersen}}, \bibinfo {author} {\bibfnamefont {T.~B.}\ \bibnamefont
			{Schr{\o}der}}, \ and\ \bibinfo {author} {\bibfnamefont {J.~C.}\ \bibnamefont
			{Dyre}},\ }\bibfield  {title} {\enquote {\bibinfo {title} {Phase diagram of
				{Kob-Andersen}-type binary {Lennard-Jones} mixtures},}\ }\href {\doibase
		10.1103/PhysRevLett.120.165501} {\bibfield  {journal} {\bibinfo  {journal}
			{Phys. Rev. Lett.}\ }\textbf {\bibinfo {volume} {120}},\ \bibinfo {pages}
		{165501} (\bibinfo {year} {2018})}\BibitemShut {NoStop}%
	\bibitem [{\citenamefont {Bell}\ \emph {et~al.}(2020)\citenamefont {Bell},
		\citenamefont {Dyre},\ and\ \citenamefont {Ingebrigtsen}}]{bel20}%
	\BibitemOpen
	\bibfield  {author} {\bibinfo {author} {\bibfnamefont {I.~H.}\ \bibnamefont
			{Bell}}, \bibinfo {author} {\bibfnamefont {J.~C.}\ \bibnamefont {Dyre}}, \
		and\ \bibinfo {author} {\bibfnamefont {T.~S.}\ \bibnamefont {Ingebrigtsen}},\
	}\bibfield  {title} {\enquote {\bibinfo {title} {Excess-entropy scaling in
				supercooled binary mixtures},}\ }\href {\doibase 10.1038/s41467-020-17948-1}
	{\bibfield  {journal} {\bibinfo  {journal} {Nat. Commun.}\ }\textbf {\bibinfo
			{volume} {11}},\ \bibinfo {pages} {4300} (\bibinfo {year}
		{2020})}\BibitemShut {NoStop}%
	\bibitem [{\citenamefont {Schr{\o}der}\ and\ \citenamefont
		{Dyre}(2020)}]{sch20}%
	\BibitemOpen
	\bibfield  {author} {\bibinfo {author} {\bibfnamefont {T.~B.}\ \bibnamefont
			{Schr{\o}der}}\ and\ \bibinfo {author} {\bibfnamefont {J.~C.}\ \bibnamefont
			{Dyre}},\ }\bibfield  {title} {\enquote {\bibinfo {title} {Solid-like
				mean-square displacementin glass-forming liquids},}\ }\href {\doibase
		10.1063/5.0004093} {\bibfield  {journal} {\bibinfo  {journal} {J. Chem.
				Phys.}\ }\textbf {\bibinfo {volume} {152}},\ \bibinfo {pages} {141101}
		(\bibinfo {year} {2020})}\BibitemShut {NoStop}%
	\bibitem [{\citenamefont {Powles}\ \emph {et~al.}(2005)\citenamefont {Powles},
		\citenamefont {Rickayzen},\ and\ \citenamefont {Heyes}}]{pow05}%
	\BibitemOpen
	\bibfield  {author} {\bibinfo {author} {\bibfnamefont {J.~G.}\ \bibnamefont
			{Powles}}, \bibinfo {author} {\bibfnamefont {G.}~\bibnamefont {Rickayzen}}, \
		and\ \bibinfo {author} {\bibfnamefont {D.~M.}\ \bibnamefont {Heyes}},\
	}\bibfield  {title} {\enquote {\bibinfo {title} {Temperatures: old, new and
				middle aged},}\ }\href@noop {} {\bibfield  {journal} {\bibinfo  {journal}
			{Mol. Phys.}\ }\textbf {\bibinfo {volume} {103}},\ \bibinfo {pages}
		{1361--1373} (\bibinfo {year} {2005})}\BibitemShut {NoStop}%
	\bibitem [{\citenamefont {Gnan}\ \emph {et~al.}(2009)\citenamefont {Gnan},
		\citenamefont {Schr{\o}der}, \citenamefont {Pedersen}, \citenamefont
		{Bailey},\ and\ \citenamefont {Dyre}}]{IV}%
	\BibitemOpen
	\bibfield  {author} {\bibinfo {author} {\bibfnamefont {N.}~\bibnamefont
			{Gnan}}, \bibinfo {author} {\bibfnamefont {T.~B.}\ \bibnamefont
			{Schr{\o}der}}, \bibinfo {author} {\bibfnamefont {U.~R.}\ \bibnamefont
			{Pedersen}}, \bibinfo {author} {\bibfnamefont {N.~P.}\ \bibnamefont
			{Bailey}}, \ and\ \bibinfo {author} {\bibfnamefont {J.~C.}\ \bibnamefont
			{Dyre}},\ }\bibfield  {title} {\enquote {\bibinfo {title} {Pressure-energy
				correlations in liquids. {IV. ``Isomorphs''} in liquid phase diagrams},}\
	}\href@noop {} {\bibfield  {journal} {\bibinfo  {journal} {J. Chem. Phys.}\
		}\textbf {\bibinfo {volume} {131}},\ \bibinfo {pages} {234504} (\bibinfo
		{year} {2009})}\BibitemShut {NoStop}%
	\bibitem [{\citenamefont {B{\o}hling}\ \emph {et~al.}(2012)\citenamefont
		{B{\o}hling}, \citenamefont {Ingebrigtsen}, \citenamefont {Grzybowski},
		\citenamefont {Paluch}, \citenamefont {Dyre},\ and\ \citenamefont
		{Schr{\o}der}}]{boh12}%
	\BibitemOpen
	\bibfield  {author} {\bibinfo {author} {\bibfnamefont {L.}~\bibnamefont
			{B{\o}hling}}, \bibinfo {author} {\bibfnamefont {T.~S.}\ \bibnamefont
			{Ingebrigtsen}}, \bibinfo {author} {\bibfnamefont {A.}~\bibnamefont
			{Grzybowski}}, \bibinfo {author} {\bibfnamefont {M.}~\bibnamefont {Paluch}},
		\bibinfo {author} {\bibfnamefont {J.~C.}\ \bibnamefont {Dyre}}, \ and\
		\bibinfo {author} {\bibfnamefont {T.~B.}\ \bibnamefont {Schr{\o}der}},\
	}\bibfield  {title} {\enquote {\bibinfo {title} {Scaling of viscous dynamics
				in simple liquids: {Theory,} simulation and experiment},}\ }\href@noop {}
	{\bibfield  {journal} {\bibinfo  {journal} {New J. Phys.}\ }\textbf {\bibinfo
			{volume} {14}},\ \bibinfo {pages} {113035} (\bibinfo {year}
		{2012})}\BibitemShut {NoStop}%
	\bibitem [{\citenamefont {Ingebrigtsen}\ and\ \citenamefont
		{Tanaka}(2015)}]{ing15}%
	\BibitemOpen
	\bibfield  {author} {\bibinfo {author} {\bibfnamefont {T.~S.}\ \bibnamefont
			{Ingebrigtsen}}\ and\ \bibinfo {author} {\bibfnamefont {H.}~\bibnamefont
			{Tanaka}},\ }\bibfield  {title} {\enquote {\bibinfo {title} {Effect of size
				polydispersity on the nature of {Lennard-Jones} liquids},}\ }\href {\doibase
		10.1021/acs.jpcb.5b02329} {\bibfield  {journal} {\bibinfo  {journal} {J.
				Phys. Chem. B}\ }\textbf {\bibinfo {volume} {119}},\ \bibinfo {pages}
		{11052--11062} (\bibinfo {year} {2015})}\BibitemShut {NoStop}%
	\bibitem [{\citenamefont {Schr{\o}der}\ \emph {et~al.}(2009)\citenamefont
		{Schr{\o}der}, \citenamefont {Bailey}, \citenamefont {Pedersen},
		\citenamefont {Gnan},\ and\ \citenamefont {Dyre}}]{III}%
	\BibitemOpen
	\bibfield  {author} {\bibinfo {author} {\bibfnamefont {T.~B.}\ \bibnamefont
			{Schr{\o}der}}, \bibinfo {author} {\bibfnamefont {N.~P.}\ \bibnamefont
			{Bailey}}, \bibinfo {author} {\bibfnamefont {U.~R.}\ \bibnamefont
			{Pedersen}}, \bibinfo {author} {\bibfnamefont {N.}~\bibnamefont {Gnan}}, \
		and\ \bibinfo {author} {\bibfnamefont {J.~C.}\ \bibnamefont {Dyre}},\
	}\bibfield  {title} {\enquote {\bibinfo {title} {Pressure-energy correlations
				in liquids. {III. Statistical} mechanics and thermodynamics of liquids with
				hidden scale invariance},}\ }\href@noop {} {\bibfield  {journal} {\bibinfo
			{journal} {J. Chem. Phys.}\ }\textbf {\bibinfo {volume} {131}},\ \bibinfo
		{pages} {234503} (\bibinfo {year} {2009})}\BibitemShut {NoStop}%
\end{thebibliography}
\end{document}